 \documentclass[aps,pre,onecolumn,superscriptaddress,nofootinbib,11pt,a4]{revtex4}

\usepackage{amsfonts}
\usepackage{amsmath}
\usepackage{graphicx}        % standard LaTeX graphics tool
 \newcommand{\rme}{{\mathrm{e}}}

\newcommand{\rmpp}{{\mathrm{p}}}
\newcommand{\rmd}{{\mathrm{d}}}
\newcommand{\rmi}{{\mathrm{i}}}

\begin{document}

%\title[New foundations of basic plasma physics on classical mechanics]{New foundations and unification of basic plasma physics by means of classical mechanics}

\title{New foundations and unification of basic plasma physics \\ by means of classical mechanics}

\author{D~F Escande, F Doveil and Yves Elskens}

 \affiliation{Aix-Marseille Universit\'{e}, CNRS, PIIM, UMR 7345, \\
% \address{Aix-Marseille Universit\'{e}, CNRS, PIIM, UMR 7345, \\
   case 321, campus Saint-J\'er\^ome, FR-13013 Marseille, France}
\email{Dominique.Escande@univ-amu.fr, fabrice.doveil@univ-amu.fr, yves.elskens@univ-amu.fr}
%\eads{\mailto{dominique.escande@univ-amu.fr},
%          \mailto{fabrice.doveil@univ-amu.fr},
%          \mailto{yves.elskens@univ-amu.fr}}

\begin{abstract}
The derivation of Debye shielding and Landau damping from the $N$-body description of plasmas requires many pages of heavy kinetic calculations in classical textbooks and is done in distinct, unrelated chapters. Using Newton's second law for the $N$-body system, we perform this derivation in a few steps with elementary calculations using standard tools of calculus, and no probabilistic setting.
Unexpectedly, Debye shielding is encountered on the way to Landau damping.
The theory is extended to accommodate a correct description of trapping or chaos due to Langmuir waves,
and to avoid the small amplitude assumption for the electrostatic potential.
Using the shielded potential, collisional transport is computed for the first time by a convergent expression including the correct calculation of deflections for all impact parameters. Shielding and collisional transport are found to be two related aspects of the repulsive deflections of electrons.
%[\today]

\par\medskip
\noindent
  PACS numbers :  \newline %
  52.20.-j	 Elementary processes in plasmas \newline
  52.35.Fp  Plasma: electrostatic waves and oscillations \newline
  45.50.-j	 Dynamics and kinematics of a particle and a system of particles \newline
  05.60.Cd  Classical transport \newline
  52.25.Fi  Transport properties \newline
  05.20.Dd  Kinetic theory \newline
  
% \pacs {52.,52.25.Dg,52.35.Fp}


\end{abstract}

\par \medskip
%  \noindent  %\mathbf{preprint submitted for publication} \\
%   \textbf{DRAFT -- private communication, not for distribution}
%  % To appear in: \textit{***}.
%  \par

\maketitle

%VVVVVVVVVVVVVVVVVVVVVVVVVVVVVVVVVVVVVVV
\section{Motivation}
\label{secIntro}
%VVVVVVVVVVVVVVVVVVVVVVVVVVVVVVVVVVVVVVV

``Don't model bulldozers with quarks''. This motto by Goldenfeld and Kadanoff \cite{GoKa} illustrates the classical wisdom that one should give up the most fundamental descriptions of physics, and use more synthetic models, when dealing with complex systems. For macroscopic classical systems, the $N$-body description by classical mechanics was deemed impossible. This led to the development of thermodynamics, of fluid mechanics, and of kinetic equations to describe various macroscopic systems made up of particles like electrons, gas atoms or molecules, stars, or microorganisms. When plasma physicists had to address the microscopic description of their state(s) of matter, they did not consider the $N$-body description by classical mechanics, but directly derived kinetic analogues of the Boltzmann equation, in particular the Vlasov equation. This trend has been the dominant one till nowadays.

However, for plasmas where transport due to short range interactions is weak, $N$-body classical mechanics yields useful results. As will be recalled in section \ref{WPDRT}, it already enabled a description of wave-particle interaction making it more intuitive, incorporating modern chaotic dynamics, and unifying particle and wave evolutions, as well as collective and finite-$N$  physics \cite{AEE,EZE,EEB,Houches}. The present paper makes an even more thorough use of $N$-body mechanics by working directly with Newton's second law for this system. It shows, in particular, that basic phenomena like Debye shielding and Landau damping can be more easily derived by avoiding kinetic and statistical calculations altogether. In particular, the new derivation of Landau damping goes first through Debye shielding, a totally unexpected fact, as classical textbooks present these concepts in different and unrelated chapters.
Furthermore, $N$-body dynamics provides an intuitive explanation of Debye shielding, showing how each particle can be shielded by all other ones, while all the plasma particles are in uninterrupted motion~: this turns out to be a mere consequence of the almost independent deflections of particles due to the Coulomb force. Finally, by using the shielded potential, the present paper provides the first calculation of collisional transport without any \textit{ad hoc} cutoff, and covering all relevant scales~: the Debye length, the inter-particle distance, and the distance of minimum approach of two electrons in a Rutherford collision. It is worth noting that the mathematical tools for the present theory were essentially available more than one century ago.

``[The] very wealth of applicability [of plasma physics] has sometimes obscured
the structure and intrinsic content of the field as a physics discipline. To put the matter a little too strongly,
what sometimes emerges from plasma introductory literature is the impression of a collection of recipes.''
This statement by Hazeltine and Waelbroeck in the preface of their book \cite{HW} may be substantiated in various ways.
Here are some elements in this line, motivating the present paper.
\begin{itemize}
\item{
First, the derivation of the Vlasov equation from first principles is painstaking (see e.g.\  \cite{Nicholson} and chapter 5 of \cite{Spohn}), and most textbooks prefer to introduce it with qualitative intuitive arguments only. Its mathematical derivation for particles interacting through the (divergent) Coulomb force is still an open problem \cite{Kie13}. This equation is hard to grasp for students, and is an obstacle for non-experts interested in kinetic aspects of plasmas.
}
\item{
The Vlasovian derivations of Landau damping do not provide the description of the corresponding evolution of particles. This description is provided in textbooks by complementary approximate mechanical models. It is worth recalling that, because of the lack of intuitive contents of Vlasovian derivations, the reality of Landau damping was fully recognized only after its experimental observation in 1964 by Malmberg and Wharton \cite{MW}, almost two decades after its prediction.
}
\item{
In principle, Vlasov equation may also be applied to velocity distributions which are positive measures
(see ch.~5 of \cite{Spohn}). This makes it applicable to non-smooth distributions (for instance two-stream ones),
but textbooks generally prefer dealing with such cases by using a fluid description of the plasma,
 at the cost of a conceptual zigzag.
}
\item{
The complete traditional derivation of Debye shielding involves the equilibrium pair correlation function
which is computed after deriving the first two equations of the BBGKY hierarchy
and truncating the cluster expansion to order 2 (see  e.g.\ ch.\ 12 of \cite{BoydSan}).
However, most textbooks prefer to introduce this shielding by adding a test particle
to a Vlasovian plasma or to a fluid one with Boltzmannian electrons.
These recipes, though efficient, are conceptual zigzags,
since they introduce a particle in descriptions resulting from the previous smoothing of plasma graininess.
The Vlasovian calculation does not reveal
how all particles shield the other ones and are also shielded by them at the same time.
The fluid calculation of shielding appeals to the ability of particles to move
and neutralize any region of excess space charge,
which makes sense if there is a macroscopic polarized Langmuir probe, but not for uniform plasmas.
Furthermore, as shown by the first two approaches, in reality the shielding of a particle depends on its velocity,
and in general the Yukawa-type contribution must be complemented with a $1/r^3$ contribution \cite{MJS,Dscree}.
}
\item{
Collisional transport is described in textbooks with two opposite points of view~: the two-body Rutherford collision picture and a mean-field approach. The two-body Rutherford collision picture describes correctly collisions for impact parameters $b \ll d$, the interparticle distance.
However, transport coefficients are then computed by an \textit{ad hoc} extension of the integrals over $b$ up to about the Debye length $\lambda_{\rm{D}} \gg d$, which involves the Coulomb logarithm as a factor with some uncertainty. The mean-field approach is based on the Balescu-Lenard equation, and describes correctly collisions for $b \gg d$. However, transport coefficients are then computed by an \textit{ad hoc} extension of the integrals over $b$ down to $\lambda_{\rm{ma}}$, the classical distance of minimum approach (much smaller than $d$), which involves again the Coulomb logarithm as a factor with some uncertainty. The agreement between the two recipes gives confidence in their result, but till now no description of collisional transport has been describing correctly the scales about $d$.
}
\end{itemize}

Therefore, to an outsider, the derivations in plasma introductory literature lack unity, and do not look as following strictly rules of inference from first principles, as do many fields of physics. The present paper contributes to following these rules and to unifying basic plasma physics. It provides new foundations for this physics, and endows it with a special status. Indeed, an old dream comes true~:
classical mechanics can genuinely describe non trivial aspects of the macroscopic dynamics of a many-body system.

%VVVVVVVVVVVVVVVVVVVVVVVVVVVVVVVVVVVVVVV
\section{Main results and paper outline}
\label{MR}
%VVVVVVVVVVVVVVVVVVVVVVVVVVVVVVVVVVVVVVV

Here are the main results of this paper and its organization~:
\begin{enumerate}
\item{\label{claim-1}
In section \ref{FEP}, by using the Fourier and Laplace transforms in a way similar to that of the Vlasovian derivation of Landau damping, a rigorous equation (Eq.~(\ref{phihat})) is derived for a linearized version of the electrostatic potential of an infinite plasma made up of the periodic replication of $N$ electrons coupled by Coulomb forces in a volume $L^3$ with a neutralizing ionic background (One Component Plasma (OCP) model \cite{Salp,Abe,BH}). This equation is of the type ${\mathcal{E}}\hat \varphi= {\mathcal{S}}$, where ${\mathcal{E}}$ is a linear operator, acting on the infinite dimensional array $\hat \varphi$ whose components are all the Doppler shifted Fourier-Laplace components of the potential. Both ${\mathcal{E}}$ and the source term ${\mathcal{S}}$ are sums over the $N$ particles. Appendix \ref{FNLEP} yields a rigorous fully nonlinear version of Eq.~(\ref{phihat})~: Eq.~(\ref{phihatNL}).
}
\item{\label{claim-2}
In section \ref{SCP}, the discrete sums in ${\mathcal{E}}$  are substituted with integrals over a smooth distribution function $f(\mathbf{r},\mathbf{v})$ close to a uniform one. Then ${\mathcal{E}}$ becomes diagonal, and the new approximate potential turns out to be the sum of the shielded Coulomb potentials of the individual particles (Eq.~(\ref{phi})). Such potentials were first computed by a kinetic approach in section II.A of Ref.\ \cite{Gasio} and later on in \cite{Bal,Rost}. Therefore, Debye shielding is computed for a single mechanical realization of the plasma.
}
\item{\label{claim-3}
In section \ref{STSP}, the discrete sums over particles of their shielded potentials are substituted with integrals over $f(\mathbf{r},\mathbf{v})$. This yields Eqs (\ref{phihatL}) and (\ref{phi0hatcg}) enabling the calculation of Langmuir waves excited by a small initial perturbation in plasmas with a possibly non-smooth $f(\mathbf{v})$ (for instance a two-stream one). For a smooth $f$, one recovers the classical Vlasovian expression including initial conditions in Landau contour calculations of Langmuir wave growth or damping, obtained by linearizing Vlasov equation and using Fourier-Laplace transform, as described in many textbooks (see for instance Refs \cite{HW,Nicholson,BoydSan}). Therefore, in these calculations, the electrostatic potential turns out to be the smoothed version of the actual shielded potential in the plasma. Sections \ref{FEP} to \ref{STSP} provide the explicit, yet very compact derivation of formulas requiring at least twenty pages in classical textbooks proceeding also explicitly from the $N$-body description. This occurs thanks to a considerable simplification of the mathematical framework with respect to textbooks, in particular because no probabilistic argument and no partial differential equation are used.
}
\item{\label{claim-4}
In section \ref{MIIDS}, Picard iteration technique (one of the standard methods to prove the existence and uniqueness of solutions to first-order equations with given initial conditions) is applied to the equation of motion of a particle $P$ due to the Coulomb forces of all other ones. It stresses now that a part of the effect on particle $P$ of another particle $P'$ is mediated by all other particles (Eq.~(\ref{rsecnAccDev2})). Indeed particle $P'$ modifies the motion of all other particles, implying that the action of the latter ones on particle $P$ is modified by particle $P'$.
}
\item{\label{claim-5}
This calculation yields the following interpretation of shielding.
At $t = 0$ consider a set of (uniformly, independently) randomly distributed particles, and especially particle $P$. At a later time $t$, the latter has deflected all particles which made a closest approach to it with a typical impact parameter $b \lesssim v_{\rm{th}} t$ where $v_{\rm{\rm{th}}}$ is the thermal velocity. This part of their global deflection due to particle $P$ reduces the number of particles inside the sphere $S(t)$ of radius $v_{\rm{th}} t$ about it. Therefore, according to Gauss' theorem, the effective charge of particle $P$ as seen out of $S(t)$ is reduced~: the charge of particle $P$ is shielded due to these deflections. This shielding effect increases with $t$, and thus with the distance to particle $P$. It becomes complete at a distance on the order of $\lambda_{\rm{D}}$. As a result, when starting from random particle positions, the typical time-scale for shielding to set in is the time for a thermal particle to cross a Debye sphere, i.e.\ $\omega_{\rmpp}^{-1}$,  where $\omega_{\rmpp}$ is the plasma frequency. Furthermore, shielding, though very fast a process, is a cooperative dynamical one, not a collective one~: it results from the accumulation of almost independent repulsive deflections with the same qualitative impact on the effective electric field of particle $P$ (if point-like ions were present, the attractive deflection of charges with opposite signs would have the same effect). So, shielding and collisional transport are two aspects of the same two-body repulsive process.
}
\item{\label{claim-6}
In section \ref{WPDRT}, in the spirit of Refs \cite{OWM,OLMSS,AEE,EEB}, to accommodate a correct description of trapping or chaos due to Langmuir waves, the set of particles is split into bulk and tail, where the bulk is the set of particles which cannot resonate with Langmuir waves. Repeating for the bulk particles the analysis leading to Eq.~(\ref{phihat}), the same equation is recovered with an additional source term due to the tail particles (Eq.~(\ref{phihatU})).
}
\item{\label{claim-7}
Using the fact that the number of tail particles is small with respect to the bulk one, and a technique introduced in Refs \cite{OWM,OLMSS}, an amplitude equation is derived for any Fourier component of the potential where tail particles provide a source term (Eq.~(\ref{eqampl})).
}
\item{\label{claim-8}
This equation, together with the equation of motion of the tail particles, enables to show that, in the linear regime, the amplitude of a Langmuir wave is ruled by Landau growth or damping, and by spontaneous emission (Eq.~(\ref{evampfinal})), a generalization to 3 dimensions of the one-dimensional result of Refs \cite{EZE,EEB}.
}
\item{\label{claim-9}
In section \ref{DSCT}, by using the shielded potential,
the trace $T_D$ of the diffusion tensor of a given particle is computed by a convergent expression
including the particle deflections for all impact parameters.
These deflections are computed by first order perturbation theory in the total electric field,
except for those due to close encounters.
The contribution to $T_D$ of the former ones is matched with that of the latter ones provided by Ref.\ \cite{Ros}.
The detailed matching procedure includes the scale of the inter-particle distance,
and is reminiscent of that in Ref.\ \cite{Hub}, without invoking the cancellation of three infinite integrals.
$T_D$ has the same expression as that in Ref.\ \cite{Ros},
except for the Coulomb logarithm which is modified by a velocity dependent quantity of order 1.
}
\item{\label{claim-10}
Appendix \ref{DA1} discusses the corrections to the ballistic approximation and the Coulomb potential.
}
\item{\label{claim-11}
Appendix \ref{FNLEP} derives the fundamental nonlinear equation for the electric potential.
}
\item{\label{claim-12}
Appendix \ref{DSmoo} discusses the smoothing procedure.
}
\end{enumerate}

%VVVVVVVVVVVVVVVVVVVVVVVVVVVVVVVVVVVVVVV
\section{Fundamental linear equation for the potential}
\label{FEP}
%VVVVVVVVVVVVVVVVVVVVVVVVVVVVVVVVVVVVVVV

This paper deals with the One Component Plasma (OCP) model \cite{Salp,Abe,BH}, which considers the plasma as infinite with spatial periodicity $L$ in three orthogonal directions with coordinates $(x,y,z)$, and made up of $N$ electrons in each elementary cube with volume $L^3$. Ions are present only as a uniform neutralizing background, enabling periodic boundary conditions. This choice is made to simplify the analysis which focuses on $\varphi(\mathbf{r})$, the potential created by the $N$ particles at any point where there is no particle. The discrete Fourier transform of $\varphi$, readily obtained from the Poisson equation, is given by $\tilde{\varphi}(\mathbf{0}) = 0$, and for $\mathbf{m} \neq \mathbf{0}$ by
\begin{equation}
  \tilde{\varphi}(\mathbf{m})
  = -\frac{e}{\epsilon_0 k_{\mathbf{m}}^2} \sum_{j \in S}
     \exp(- \rmi \mathbf{k}_{\mathbf{m}} \cdot \mathbf{r}_j),
\label{phitildetotM}
\end{equation}
where $-e$ is the electron charge, $\epsilon_0$ is the vacuum permittivity, $\mathbf{r}_j$ is the position of particle $j$,
$S = \{ 1, \ldots N \}$, $\tilde{\varphi}(\mathbf{m})
= \int \varphi(\mathbf{r}) \exp(- \rmi \mathbf{k}_{\mathbf{m}} \cdot \mathbf{r}) \, \rmd^3 \mathbf{r}$,
with $\mathbf{m} = (m_x,m_y,m_z)$ a vector with three integer components running from $- \infty$ to $+ \infty$, $\mathbf{k}_{\mathbf{m}} = \frac{2 \pi}{L} \, \mathbf{m}$, and $k_{\mathbf{m}} = \|\mathbf{k}_{\mathbf{m}}\|$. Reciprocally,
\begin{equation}
\varphi(\mathbf{r}) = \frac{1}{L^3}\sum_{\mathbf{m}} \tilde{\varphi} (\mathbf{m}) \exp(\rmi \mathbf{k}_{\mathbf{m}} \cdot \mathbf{r}) .
\label{phiInv}
\end{equation}

The dynamics of particle $l$ follows Newton's equation
\begin{equation}
  \ddot{\mathbf{r}}_l
  = \frac{e}{m_\rme} \nabla \varphi_l(\mathbf{r}_l),
\label{rsectot}
\end{equation}
with $m_\rme$ the electron mass, and $\varphi_l$ the electrostatic potential acting on particle $l$, i.e.\ the one created by all other particles and by
the background charge.
Its Fourier transform is given by Eq.\ (\ref{phitildetotM})
with the restriction $j \neq l$.
Let
\begin{equation}
  \mathbf{r}_l^{(0)}
  = \mathbf{r}_{l0} + \mathbf{v}_{l} t
\label{rl0}
\end{equation}
be a ballistic approximation to the motion of particle $l$, and let $\delta \mathbf{r}_l = \mathbf{r}_l - \mathbf{r}_l^{(0)}$. In the following, we consider two instances of the ballistic approximation~: the one where $\mathbf{r}_{l0}$ and $\mathbf{v}_{l}$ are respectively the initial position and velocity of particle $l$, and the one where they are slightly shifted from these values by low amplitude Langmuir waves.
Until the end of section \ref{DSLD}, we consider cases where all the $\delta \mathbf{r}_l$'s are small.
So we approximate $\tilde{\varphi}_l(\mathbf{m})$ by its expansion
to first order in the $\delta \mathbf{r}_l$'s (Approximation 1, discussed in Appendix \ref{DA1})
\begin{equation}
\tilde{\phi}_l (\mathbf{m}) = \sum_{j \in S;j \neq l} \delta \tilde{\phi}_{j} (\mathbf{m}),
\label{phitildn}
\end{equation}
with the contribution of particle $j$ to the potential reading
\begin{equation}
\delta \tilde{\phi}_{j} (\mathbf{m}) = -\frac{e}{\epsilon_0 k_{\mathbf{m}}^2} \exp(- \rmi \mathbf{k}_{\mathbf{m}} \cdot \mathbf{r}_{j}^{(0)})(1 - \rmi \mathbf{k}_{\mathbf{m}} \cdot \delta \mathbf{r}_j).
\label{phitildnj}
\end{equation}

We further consider $\varphi$ to be small, and the $\delta \mathbf{r}_l$'s to be of the order of $\varphi$ (Approximation 2). At lowest order, the particles dynamics defined by Eq.\ (\ref{rsectot}) is given by
\begin{equation}
  \delta \ddot{\mathbf{r}}_l
  = \frac{\rmi e}{L^3 m_\rme} \sum_{\mathbf{n}} \mathbf{k}_{\mathbf{n}} \
    \tilde{\phi}_l(\mathbf{n}) \exp(\rmi \mathbf{k}_{\mathbf{n}} \cdot \mathbf{r}_l^{(0)}).
\label{delrsec}
\end{equation}

We denote with a caret the time Laplace transform which maps a function $ f(t)$
to $\widehat{f}(\omega) = \int_0^{\infty}  f(t) \exp(\rmi \omega t) \rmd t$
(with $\omega$ complex).
In particular, we first define the ballistic approximation $\widehat{\tilde{\phi}}_l^{(0)}$
to the Laplace transform of $\tilde{\phi}_l (\mathbf{m})$~:
it is computed from Eqs (\ref{phitildn}) and (\ref{phitildnj})
on setting $\delta \mathbf{r}_j = \delta \mathbf{r}_j(0) + \delta \dot{\mathbf{r}}_j(0) t$
for all $j$'s in the latter,
\begin{equation}
  \widehat{\tilde{\phi}}_l^{(0)}(\mathbf{m},\omega) =
  \sum_{j \in S;j \neq l} \delta \widehat{\tilde{\phi}}_j^{(0)}(\mathbf{m},\omega)
  ,
\label{phi0hat}
\end{equation}
where
\begin{equation}
  \delta \widehat{\tilde{\phi}}_j^{(0)}(\mathbf{m},\omega)
  = - \frac{\rmi e}{\epsilon_0 k_{\mathbf{m}}^2}
      \frac{\exp[- \rmi \mathbf{k}_{\mathbf{m}}  \cdot (\mathbf{r}_{j0} + \delta \mathbf{r}_j(0))]}
             {\omega -\mathbf{k}_{\mathbf{m}}  \cdot (\mathbf{v}_j + \delta \dot{\mathbf{r}}_j(0))}
\label{phij0hat}
\end{equation}
is the ballistic contribution of particle $j$ to the total potential.

The Laplace transform of Eq.\ (\ref{delrsec}) is
\begin{equation}
%\fl
  \omega^2 \delta \widehat{\mathbf{r}}_l(\omega)
  = - \frac{\rmi e}{L^3 m_\rme} \sum_{\mathbf{n}}
                \mathbf{k}_{\mathbf{n}} \exp(\rmi \mathbf{k}_{\mathbf{n}} \cdot \mathbf{r}_{l0})
                  \ \widehat{\tilde{\phi}}_l(\mathbf{n},\omega + \omega_{\mathbf{n},l})
     + \rmi \omega \delta \mathbf{r}_l(0) - \delta \dot{\mathbf{r}}_l(0) ,
\label{rLapl}
\end{equation}
where $\omega_{\mathbf{n},l} = \mathbf{k}_{\mathbf{n}} \cdot \mathbf{v}_{l}$
comes from the time dependence of $\mathbf{r}_l^{(0)}$ in the exponent of Eq.\ (\ref{delrsec}).
The Laplace transform of Eqs\ (\ref{phitildn})-(\ref{phitildnj}), with the actual $\delta \mathbf{r}_j(t)$, then yields
\begin{equation}
%\fl
 k_{\mathbf{m}}^2\widehat{\tilde{\phi}}_l(\mathbf{m},\omega) = k_{\mathbf{m}}^2 \widehat{\tilde{\phi}}_l^{(00)}(\mathbf{m},\omega) + \frac{\rmi e}{\epsilon_0} \sum_{j \in S;j \neq l} \exp(- \rmi \mathbf{k}_{\mathbf{m}} \cdot \mathbf{r}_{j0}) \ \mathbf{k}_{\mathbf{m}} \cdot \delta \widehat{\mathbf{r}}_j(\omega - \omega_{\mathbf{m},j}),
\label{phihatn}
\end{equation}
where $\omega_{\mathbf{m},j}$ comes from the $\mathbf{r}_j^{(0)}$ in Eq.\  (\ref{phitildnj})~;
$\widehat{\tilde{\phi}}_l^{(00)}(\mathbf{m},\omega) $ is
$\widehat{\tilde{\phi}}_l^{(0)}(\mathbf{m},\omega)$ computed with
$\delta \mathbf{r}_j(0) = \delta \dot{\mathbf{r}}_j(0)  = 0$ for all $j$'s.
On substituting the $\delta \widehat{\mathbf{r}}_j$'s with their expression,
Eq.\ (\ref{phihatn}) becomes
\begin{eqnarray}
%\fl  
&&
  k_{\mathbf{m}}^2 \widehat{\tilde{\phi}}_l(\mathbf{m},\omega)
%  \nonumber \\ &&
  - \frac{e^2}{ L^3 m_\rme \epsilon_0}
      \sum_{\mathbf{n}} \mathbf{k}_{\mathbf{m}} \cdot \mathbf{k}_{\mathbf{n}}
     \   \sum_{j \in S;j \neq l} \frac{\widehat{\tilde{\phi}}_j(\mathbf{n},\omega + \omega_{\mathbf{n},j} - \omega_{\mathbf{m},j})}{(\omega - \omega_{\mathbf{m},j})^2} \exp[\rmi (\mathbf{k}_{\mathbf{n}}-\mathbf{k}_{\mathbf{m}}) \cdot \mathbf{r}_{j0}]
 \nonumber \\
%\fl 
&=&  k_{\mathbf{m}}^2 \widehat{\tilde{\phi}}_l^{(0)}(\mathbf{m},\omega) .
\label{phihatnf}
\end{eqnarray}

Summing Eq.\ (\ref{phihatnf}) over $l = 1,... N$ and dividing by $N-1$  yields
\begin{eqnarray}
%\fl  
&&
   k_{\mathbf{m}}^2\widehat{\tilde{\phi}}(\mathbf{m},\omega)
 - \frac{e^2}{ L^3 m_\rme \epsilon_0}
 \sum_{\mathbf{n}} \mathbf{k}_{\mathbf{m}} \cdot \mathbf{k}_{\mathbf{n}}
\ \sum_{j \in S} \frac{\widehat{\tilde{\phi}}(\mathbf{n},\omega + \omega_{\mathbf{n},j} - \omega_{\mathbf{m},j})}{(\omega - \omega_{\mathbf{m},j})^2} \exp[\rmi (\mathbf{k}_{\mathbf{n}}-\mathbf{k}_{\mathbf{m}}) \cdot \mathbf{r}_{j0}]
 \nonumber \\
%\fl 
& = &  k_{\mathbf{m}}^2 \widehat{\tilde{\phi}}^{(0)}(\mathbf{m},\omega) ,
\label{phihat}
\end{eqnarray}
where $\widehat{\tilde{\phi}}(\mathbf{m},\omega)$ and $\widehat{\tilde{\phi}}^{(0)}(\mathbf{m},\omega) $ are respectively $\widehat{\tilde{\phi}}_l (\mathbf{m},\omega)$ and $\widehat{\tilde{\phi}}_l^{(0)}(\mathbf{m},\omega) $ complemented with the missing $l$-th term. Equation (\ref{phihat}) is the fundamental linear equation of this paper. \textit{This fundamental linear equation is of the type ${\mathcal{E}}\widehat{\tilde{ \phi}}=$ source term}, where ${\mathcal{E}}$ is a linear operator, acting on the infinite dimensional array whose components are all the Doppler shifted $\widehat{\tilde{\phi}}(\mathbf{m},\omega)$'s.

A fully nonlinear and rigorous version of the fundamental linear equation is provided in Appendix \ref{FNLEP}~: Eq.~(\ref{phihatNL}). Its linearization provides Eq.~(\ref{phihat}), which endows it with a status analogous to the linearized version of the nonlinear Vlasov-Poisson system of equations. Since the nonlinear version is not used in this paper, for simplicity we derived here the linearized version only.

%VVVVVVVVVVVVVVVVVVVVVVVVVVVVVVVVVVVVVVV
\section{Debye shielding, Langmuir waves and Landau damping}
\label{DSLD}
%VVVVVVVVVVVVVVVVVVVVVVVVVVVVVVVVVVVVVVV

%VVVVVVVVVVVVVVVVVVVVVVVVVVVVVVVVVVVVVVV
\subsection{Shielded Coulomb potential}
\label{SCP}
%VVVVVVVVVVVVVVVVVVVVVVVVVVVVVVVVVVVVVVV

We introduce a smooth function $f(\mathbf{r},\mathbf{v})$, the \textit{smoothed} position and velocity distribution
function at $t=0$ such that the distribution
\begin{equation}
    \sum_{l \in S} \bullet
    =
    \iint  \bullet  f(\mathbf{r},\mathbf{v}) \rmd^3 \mathbf{r} \, \rmd^3 \mathbf{v} + W(\bullet),
\label{fxv}
\end{equation}
where the distribution $W$ yields a negligible contribution when applied to space dependent functions which evolve slowly on the scale of the inter-particle distance~; there the spatial integration is performed over the elementary cube with volume $L^3$, and the velocity integration runs over all velocities.

On replacing the discrete sums over particles with integrals over the smooth distribution function $f(\mathbf{r},\mathbf{v})$ (Approximation 3 discussed in Appendix \ref{DSmoo}), Eq.\ (\ref{phihat}) becomes
\begin{eqnarray}
%\fl 
&&
  k_{\mathbf{m}}^2 \widehat{\tilde{\Phi}}(\mathbf{m},\omega)
 \nonumber \\
%\fl  
&=& k_{\mathbf{m}}^2 \widehat{\tilde{\phi}}^{(0)}(\mathbf{m},\omega)
    + \frac{e^2}{L^3 m_\rme \epsilon_0} \sum_{\mathbf{n}}
        \mathbf{k}_{\mathbf{m}} \cdot \mathbf{k}_{\mathbf{n}}
    \int  \frac{\widehat{\tilde{\Phi}}(\mathbf{n},
\omega + (\mathbf{k}_{\mathbf{n}} - \mathbf{k}_{\mathbf{m}}) \cdot \mathbf{v})}
               {(\omega -\mathbf{k}_{\mathbf{m}} \cdot \mathbf{v})^2}
  \tilde{f}({\mathbf{n}} - {\mathbf{m}},\mathbf{v}) \ \rmd^3 \mathbf{v},
  \nonumber \\
%\fl 
&&
\label{phihatcg}
\end{eqnarray}
where $\widehat{\tilde{\Phi}}$ is the smoothed version of $\widehat{\tilde{\varphi}}$ resulting from Approximations 1 to 3, and $\tilde{f}$ is the spatial Fourier transform of $f$. We further assume the initial distribution $f$ to be a spatially uniform distribution function $f_0(\mathbf{v})$ plus a small perturbation of the order of $\Phi$
(in agreement with Approximation 2).
Then operator $\mathcal{E}$ becomes diagonal with respect to both $\mathbf{m}$ and $\omega$ (a complex quantity).
Linearizing Eq.\ (\ref{phihatcg}) for $\widehat{\tilde{\Phi}}$
amounts to replacing $\tilde f$ with its $\Phi$-independent part, so that
\begin{equation}
  \epsilon(\mathbf{m},\omega) \widehat{\tilde{\Phi}}(\mathbf{m},\omega)
  = \widehat{\tilde{\phi}}^{(0)}(\mathbf{m},\omega),
\label{phihatL}
\end{equation}
where
\begin{equation}
  \epsilon(\mathbf{m},\omega)
  = 1 - \frac{e^2}{L^3 m_\rme \epsilon_0}
     \int \frac{f_0(\mathbf{v}) }{(\omega -\mathbf{k}_{\mathbf{m}}  \cdot \mathbf{v})^2} \ \rmd^3 \mathbf{v}.
\label{eps}
\end{equation}
This shows that the smoothed self-consistent potential $\widehat{\tilde{\Phi}}$
is determined by the response function $\epsilon(\mathbf{m},\omega)$,
viz.\  the classical plasma dielectric function.
A first check of this can be obtained for a cold plasma~:
then $\epsilon(\mathbf{m},\omega) = 1 - {\omega_{\rmpp}^2}/{\omega^2}$,
where $\omega_{\rmpp} = [(e^2 n)/(m_\rme \epsilon_0)]^{1/2}$ is the plasma frequency
($n = N/L^3 = L^{-3} \iint f(\mathbf{r},\mathbf{v})\, \rmd^3 \mathbf{r} \, \rmd^3 \mathbf{v}$
is the plasma density).
The classical expression involving $\partial f_0 / \partial \mathbf{v}$
obtains by a mere integration by parts if $f_0$ is differentiable.

As a result of Eq.\ (\ref{phi0hat}),
the part of $\widehat{\tilde{\Phi}}(\mathbf{m},\omega)$ generated by particle $j$
is $\delta \widehat{\tilde{\Phi}}_j(\mathbf{m},\omega)
= \delta \widehat{\tilde{\phi}}_j^{(0)}(\mathbf{m},\omega)/\epsilon(\mathbf{m},\omega)$.
By inverse Fourier-Laplace transform, after some transient discussed later,
the potential due to particle $j$ becomes the shielded Coulomb potential \cite{Gasio,Bal,Rost}
\begin{equation}
  \delta \Phi_j (\mathbf{r})
  = \delta \Phi(\mathbf{r} - \mathbf{r}_j(0) - \dot{\mathbf{r}}_j(0) t,\dot{\mathbf{r}}_j(0)),
\label{phij}
\end{equation}
where
\begin{equation}
  \delta \Phi (\mathbf{r},\mathbf{v})
  = - \frac{e}{L^3 \epsilon_0} \sum_{{\mathbf{m}} \neq {\mathbf{0}}}
      \frac{\exp(\rmi \mathbf{k}_{\mathbf{m}} \cdot \mathbf{r})}
           { k_{\mathbf{m}}^2 \, \epsilon(\mathbf{m},\mathbf{k}_{\mathbf{m}} \cdot \mathbf{v} + \rmi \varepsilon)}
\label{phi}
\end{equation}
with the usual $\rmi \varepsilon$ prescription resulting from inverting the Laplace transform as
the integral in Eq\ (\ref{eps}) is undefined for the real-valued
$\omega = \mathbf{k}_{\mathbf{m}} \cdot \mathbf{v}$.
Therefore, after this transient, \textit{the dominant contribution to the full potential in the plasma turns out to be
the sum of the shielded Coulomb potentials of individual particles}
located at their ballistic positions computed with their initial position and velocity.

Let $\lambda_{\rm{D}} = [(\epsilon_0 k_{\rm{B}} T)/(n e^2)]^{1/2}
     = [k_{\rm{B}} T / m_\rme]^{1/2} \omega_{\rmpp}^{-1}$
be the Debye length, where $k_{\rm{B}}$ is the Boltzmann constant and $T$ the temperature.
The wavenumbers resolving scale $\| \mathbf{r}\|$ are such that $k_{\mathbf{m}} \|\mathbf{r}\| \gtrsim 1$.
Shielding involves scales on the order of $\lambda_{\rm{D}}$.
The transient is given by the zeros of $\epsilon(\mathbf{m},\omega)$.
For shielding scales, these zeros correspond to a strong damping over time scales on the order of the plasma period.
Therefore, the transient is damped after such a period, as estimated in statement \ref{MR}.(\ref{claim-5}).
For scales much larger than $\lambda_{\rm{D}}$,
the damping is small, and particles excite weakly damped Langmuir waves too.

If $\|\mathbf{r}\| \ll \lambda_{\rm{D}}$, the corresponding wavenumbers are such
that $k_{\mathbf{m}} \lambda_{\rm{D}} \gg 1$.
Therefore, there is no shielding for $\|\mathbf{r}\| \ll \lambda_{\rm{D}}$,
since $\epsilon(\mathbf{m},\mathbf{k}_{\mathbf{m}} \cdot \mathbf{v}) - 1
  \simeq - [v_{\mathrm{th}} / (\lambda_{\rm{D}} \, \mathbf{k}_{\mathbf{m}} \cdot \mathbf{v})]^2
  \approx - (k_{\mathbf{m}} \lambda_{\rm{D}})^{-2}$
  where $v_{\mathrm{th}} = \lambda_{\rm{D}} \, \omega_{\rmpp}$ is the thermal velocity.

%VVVVVVVVVVVVVVVVVVVVVVVVVVVVVVVVVVVVVVV
\subsection{Langmuir waves and Landau damping}
\label{STSP}
%VVVVVVVVVVVVVVVVVVVVVVVVVVVVVVVVVVVVVVV

We now apply the smoothing using distribution function $f$ to $\widehat{\tilde{\phi}}^{(0)}(\mathbf{m},\omega)$ too in Eq.\ (\ref{phihatL}) (Approximation 4).
On neglecting $\delta {\mathbf{r}}_{j}$ to lowest order in Eq.\ (\ref{phitildnj}), this yields
a  $\tilde{\Phi}^{(0)}(\mathbf{m})$
whose Laplace transform is
\begin{equation}
  \widehat{\tilde{\Phi}}^{(0)}(\mathbf{m},\omega)
  = - \frac{\rmi e}{\epsilon_0 k_{\mathbf{m}}^2} \int
    \frac{\tilde{f}(\mathbf{m},\mathbf{v})}
         {\omega -\mathbf{k}_{\mathbf{m}} \cdot \mathbf{v}} \
     \rmd^3 \mathbf{v}.
\label{phi0hatcg}
\end{equation}
This shows that, whenever $f$ is differentiable in $\mathbf{v}$, this second smoothing makes Eq.\ (\ref{phihatL}) \textit{to become the expression including initial conditions in Landau contour calculations of Langmuir wave growth or damping}, usually obtained by linearizing Vlasov equation and using Fourier-Laplace transform, as described in many textbooks.

However, since Eqs (\ref{phihatL}) and (\ref{phi0hatcg}) do not involve derivatives of $f$, they also enable computing Langmuir waves induced by an initial perturbation in the case of a non differentiable $f$ (for instance a two-stream one). In all these calculations, $\widehat{\tilde{\Phi}}^{(0)}(\mathbf{m},\omega)$ turns out to be the smoothed version of the actual shielded potential in the plasma.

It is interesting to compare the above derivation with that used by classical textbooks
when they start with the $N$-body description to derive both Debye shielding
and the combination of Eqs\ (\ref{phihatL}) and (\ref{phi0hatcg}).
Debye shielding is exhibited in the equilibrium pair correlation function computed
after deriving the first two equations of the BBGKY hierarchy (see  e.g.\ chapter 12 of \cite{BoydSan}).
The combination of Eqs\ (\ref{phihatL}) and (\ref{phi0hatcg}) is obtained independently
by linearizing Vlasov equation about a uniform velocity distribution function,
and by using the Fourier-Laplace transform.
A prerequisite is the derivation of Vlasov equation by two main fundamental approaches~:
a mean-field derivation \cite{Spohn}, or the BBGKY hierarchy that involves statistical arguments
starting with the Liouville equation (see e.g.\  \cite{Nicholson}).
In contrast with the latter, the present derivation performs the Laplace transform in time of the linearized dynamics of a \textit{single realization} of the $N$-body system.
This yields Eq.\ (\ref{phihat}) which keeps the full graininess of the system.
A first smoothing involving a velocity distribution function yields Eqs (\ref{phij})-(\ref{phi}),
and a second one yields Eq.\ (\ref{phi0hatcg}) combined with Eq.\ (\ref{phihatL}).
This provides a much shorter connection between these equations and the underlying $N$-body problem.
In this derivation, the smoothed velocity distribution is introduced after particle dynamics has been taken into account,
and not before, as occurs when kinetic equations are used.
This avoids addressing the issues of the exact definition of the smoothed distribution for a given realization of the plasma,
and of the uncertainty as to the way the smoothed dynamics departs from the actual $N$-body one \cite{Spohn}.

%VVVVVVVVVVVVVVVVVVVVVVVVVVVVVVVVVVVVVVV
\subsection{Mediated interactions imply Debye shielding}
\label{MIIDS}
%VVVVVVVVVVVVVVVVVVVVVVVVVVVVVVVVVVVVVVV

In the above derivation of Debye shielding, using the Laplace transform of the particle positions does not provide an intuitive picture of this effect. We now show that such a picture can be obtained directly from the mechanical description of microscopic dynamics with the full OCP Coulomb potential of Eq.\ (\ref{phitildetotM}). To compute the dynamics, we use Picard iteration technique.
From Eq.\ (\ref{rsectot}), $\mathbf{r}_l^{(n)}$, the $n$-th iterate for $\mathbf{r}_l$, is computed from
\begin{equation}
  \ddot{\mathbf{r}}^{(n)}_l
  = \frac{e}{m_\rme} \nabla \varphi_l^{(n-1)}(\mathbf{r}^{(n-1)}_l),
\label{rsecn}
\end{equation}
where $\varphi_l^{(n-1)}$ is computed by the inverse Fourier transform of Eq.\ (\ref{phitildetotM}) with the $\mathbf{r}_j$'s substituted with the $\mathbf{r}_j^{(n-1)}$'s. The iteration starts with the ballistic approximation of the dynamics defined by Eq.\ (\ref{rl0}), and the actual orbit of Eq.\ (\ref{rsectot}) corresponds to $n \rightarrow \infty$. Let $\delta \mathbf{r}_l^{(n)} = \mathbf{r}_l^{(n)} - \mathbf{r}_l^{(0)}$ be the mismatch of the position of particle $l$ with respect to the ballistic one at the $n$-th iterate. It is convenient to write Eq.\ (\ref{rsecn}) as
$ % \begin{equation}
  \delta \ddot{\mathbf{r}}^{(n)}_l
  = \sum_{j \in S;j \neq l} \delta \ddot{\mathbf{r}}^{(n)}_{lj}
$, %\label{rsecnAccl}
% \end{equation}
with
\begin{equation}
  \delta \ddot{\mathbf{r}}^{(n)}_{lj}
  = \mathbf{a}_{\rm{C}}(\mathbf{r}_l^{(n-1)}-\mathbf{r}_j^{(n-1)})
\label{rsecnAcclj}
\end{equation}
 and
\begin{equation}
  \mathbf{a}_{\rm{C}}(\mathbf{r})
  =
  \frac{\rmi e^2}{\epsilon_0 m_\rme L^3}
  \sum_{{\mathbf{m}} \neq {\mathbf{0}}}
      k_{\mathbf{m}}^{-2} \, \mathbf{k}_{\mathbf{m}}
      \exp(\rmi \mathbf{k}_{\mathbf{m}} \cdot \mathbf{r}).
\label{phiCbPer}
\end{equation}
Let $\delta {\mathbf{r}}^{(n)}_{lj}
    = \int_0^t \int_0^{t'} \delta \ddot {\mathbf{r}}^{(n)}_{lj} (t'') \, \rmd t'' \rmd t'$.
For $n \geq 2$, one finds
\begin{equation}
  \delta \ddot{\mathbf{r}}^{(n)}_l
  = \sum_{j \in S;j \neq l}
       [(\delta \ddot{\mathbf{r}}^{(1)}_{lj} + M_{lj}^{(n-1)})
        + 2 \nabla \mathbf{a}_{\rm{C}}(\mathbf{r}_l^{(0)}-\mathbf{r}_j^{(0)}) \cdot \delta \mathbf{r}_{lj}^{(n-1)}] + O(a^3),
\label{rsecnAccDev2}
\end{equation}
where $a$ is the order of magnitude of the total Coulombian acceleration, and
\begin{equation}
  M_{lj}^{(n-1)}
  = \nabla \mathbf{a}_{\rm{C}}(\mathbf{r}_l^{(0)}-\mathbf{r}_j^{(0)})
      \cdot \sum_{i \in S; i \neq l,j} (\delta \mathbf{r}_{li}^{(n-1)} - \delta \mathbf{r}_{ji}^{(n-1)})
\label{Mlj}
\end{equation}
is the modification to the bare Coulomb acceleration of particle $j$ on particle $l$ due to the following process~:
particle $j$ modifies the position of all other particles, so that the action of the latter ones on particle $l$ is modified by particle $j$. Therefore $M_{lj}^{(n-1)} $ \textit{is the acceleration of particle $l$ due to particle $j$ mediated by all other particles}. The last term in the bracket in Eq.\ (\ref{rsecnAccDev2}) accounts for the fact that both particles $j$ and $l$ are shifted with respect to their ballistic positions.

Since the shielded potential of the previous paragraph was found by first order perturbation theory, it is felt in the acceleration of particles computed to second order. This acceleration is provided by Eq.\ (\ref{rsecnAccDev2}) for $n=2$. Therefore its term in brackets is the shielded acceleration of particle $l$ due to particle $j$. As a result, though the summation runs over all particles, its effective part is only due to particles $j$ typically inside the Debye sphere (with radius $\lambda_{\rm{D}}$) about particle $l$. Starting from the third iterate of the Picard scheme, the effective part of the summation in Eq.\ (\ref{rsecnAccDev2}) ranges inside this Debye sphere, since the $\delta \mathbf{r}_{lj}^{(n-1)}$'s are then computed with a shielded acceleration. This approach clarifies the mechanical background of the calculation of shielding using the equilibrium pair correlation function which shows shielding to result from the correlation of two particles occurring through the action of all the other ones (see e.g.\ section 12.3 of \cite{BoydSan}). The preceding calculation yields the interpretation of shielding given in statement \ref{MR}.(\ref{claim-5}).

%TTTTTTTTTTTTTTTTTTTTTTTTTTTTTTTTTTTTTT
\section{Wave-particle dynamics}
\label{WPDRT}
%TTTTTTTTTTTTTTTTTTTTTTTTTTTTTTTTTTTTTT

Section \ref{STSP} enables the calculation of Langmuir waves excited by a given initial perturbation. To describe Langmuir waves with discrete particles, we consider that the $\mathbf{r}_{l0}$'s are random, and we allow for non zero $\delta \mathbf{r}_j(0)$'s and $\delta \dot{\mathbf{r}}_j(0)$'s for the $\delta \mathbf{r}_j$'s in Eq.~(\ref{phitildnj}).
Therefore, in the formulas of section \ref{DSLD}, the $\mathbf{r}_{j0}$'s and $\mathbf{v}_j$'s
are slightly shifted with respect to the initial $\mathbf{r}_j(0)$'s and $\dot{\mathbf{r}}_j(0)$'s
due to Langmuir waves.

Up to this point, we described Langmuir waves by a fully linear theory. We now generalize the analysis of section \ref{FEP} to afford the description of nonlinear effects in wave-particle dynamics. Indeed, resonant particles may experience trapping or chaotic dynamics, which imply $\mathbf{k}_{\mathbf{m}} \cdot \delta \mathbf{r}_l$'s of the order of $2 \pi$ or larger for wave $\mathbf{k}_{\mathbf{m}}$'s. To describe such a dynamics, it is not appropriate to expand $\phi$ as was done in Eqs~(\ref{phitildn})-(\ref{phitildnj}) for such particles. However, this expansion may still be justified for non resonant particles over times where trapping and chaos show up for resonant ones. In order to keep the capability to describe the latter effects, we now split the set of $N$ particles into bulk and tail, in the spirit of Refs \cite{OWM,OLMSS,AEE,EZE,EEB}. The bulk is defined as the set of particles which are not resonant with Langmuir waves. We then perform the analysis of section \ref{FEP} for the $N_{\mathrm{bulk}}$ particles, while keeping the exact contribution of the $N_{\mathrm{tail}}$ particles to the electrostatic potential. To this end, we number the tail particles from 1 to $N_{\mathrm{tail}}$, the bulk ones from $N_{\mathrm{tail}}+1$ to $N = N_{\mathrm{bulk}} + N_{\mathrm{tail}}$, and we call these respective sets of integer $S_{\mathrm{tail}}$ and $S_{\mathrm{bulk}}$. For $l \in S_{\mathrm{bulk}}$, we now substitute Eq.~(\ref{phitildn}) with
\begin{equation}
\tilde{\phi}_l (\mathbf{m}) = \frac{N_{\mathrm{bulk}}  - 1}{N_{\mathrm{bulk}}} U(\mathbf{m})
+ \sum_{j \in S_{\mathrm{bulk}};j \neq l} \ \delta \tilde{\phi}_{j} (\mathbf{m}) ,
\label{phitildnT}
\end{equation}
where
\begin{equation}
U(\mathbf{m}) = -\frac{e N_{\mathrm{bulk}}}{\epsilon_0 k_{\mathbf{m}}^2 (N_{\mathrm{bulk}}-1)} \sum_{j \in S_{\mathrm{tail}}} \exp(- \rmi \mathbf{k}_{\mathbf{m}} \cdot \mathbf{r}_j).
\label{U}
\end{equation}
In the r.h.s.\ of Eq.~(\ref{phitildnT}), the first term vanishes if $N_{\mathrm{tail}} = 0$. We now perform the calculation of section \ref{FEP} on substituting the previous summations with index running from 1 to $N$ by ones where the index runs over $S_{\mathrm{bulk}}$, while keeping the exclusion of $j=l$ where indicated.
The previous division by $N-1$ preceding Eq.~(\ref{phihat}) is now a division by $ N_{\mathrm{bulk}} - 1$. This yields
\begin{eqnarray}
%\fl  
&&
   k_{\mathbf{m}}^2\widehat{\tilde{\phi}}(\mathbf{m},\omega)
  - \frac{e^2}{ L^3 m_\rme \epsilon_0}
 \sum_{\mathbf{n}} \mathbf{k}_{\mathbf{m}} \cdot \mathbf{k}_{\mathbf{n}}
\ \sum_{j \in S} \frac{\widehat{\tilde{\phi}}(\mathbf{n},\omega + \omega_{\mathbf{n},j} - \omega_{\mathbf{m},j})}{(\omega - \omega_{\mathbf{m},j})^2}
          \exp[\rmi (\mathbf{k}_{\mathbf{n}}-\mathbf{k}_{\mathbf{m}}) \cdot \mathbf{r}_{j0}]
\nonumber\\
%\fl  
& = &  k_{\mathbf{m}}^2 \widehat{\tilde{\phi}}^{(0)}(\mathbf{m},\omega)
           + k_{\mathbf{m}}^2 \hat{U}(\mathbf{m},\omega),
\label{phihatU}
\end{eqnarray}
where $\hat{U}(\mathbf{m},\omega)$ is the Laplace transform of $U(\mathbf{m}, t)$.
Then Eq.~(\ref{phihatL}) becomes
\begin{equation}
  \epsilon(\mathbf{m},\omega) \widehat{\tilde{\Phi}}(\mathbf{m},\omega)
  = \widehat{\tilde{\phi}}^{(0)}(\mathbf{m},\omega) + \hat{U}(\mathbf{m},\omega).
\label{phihatLU}
\end{equation}

Let $\tilde{\Phi}(\mathbf{m},t)$ be the inverse Laplace transform
of  $\widehat{\tilde{\Phi}}(\mathbf{m},\omega)$,
$\widehat{\tilde{\Phi}}_{\mathrm{bulk}}(\mathbf{m},\omega)$ be the solution of Eq.~(\ref{phihatL}) computed for the bulk particles, and $\tilde{\Phi}_{\mathrm{bulk}}(\mathbf{m},t)$ be its inverse Laplace transform. We now derive an amplitude equation for $\tilde{\Phi}(\mathbf{m},t)$ in a way similar to Refs \cite{OWM,OLMSS}.
Let $\omega_{\mathbf{m}}$ be such that $\epsilon(\mathbf{m},\omega_{\mathbf{m}}) = 0$~;
because of the definition of the bulk, this frequency is real.
Then $\tilde{\Phi}_{\mathrm{bulk}}(\mathbf{m},t) = A \exp(-\rmi \omega_{\mathbf{m}} t)$,
where $A$ is a constant, and
\begin{equation}
  \widehat{\tilde{\phi}}^{(0)}(\mathbf{m},\omega) = \frac{\rmi A}{\omega - \omega_{\mathbf{m}}},
\label{phi0A}
\end{equation}
according to Eq.~(\ref{phihatL}).

Let $g(\mathbf{m},t) = \tilde{\Phi}( \mathbf{m},t) / \tilde{\Phi}_{\mathrm{bulk}}(\mathbf{m},t)$.
Therefore $\widehat{\tilde{\Phi}}(\mathbf{m},\omega) = A\, \hat{g}(\omega - \omega_{\mathbf{m}})$,
which together with Eqs~(\ref{phihatLU}) and (\ref{phi0A}) yields
\begin{equation}
  A \, \epsilon(\mathbf{m},\omega_{\mathbf{m}} + \omega') \,
      [\hat{g}(\mathbf{m},\omega') - \frac{\rmi}{\omega'}]
  = \hat{U}(\mathbf{m},\omega_{\mathbf{m}} + \omega'),
\label{eqf}
\end{equation}
where $\omega' = \omega - \omega_{\mathbf{m}}$. If $ N_{\mathrm{tail}} \ll N_{\mathrm{bulk}}$, $g(\mathbf{m},t)$ is a slowly evolving amplitude, and the support of $\hat{g}(\mathbf{m},\omega)$ is narrow about zero. This justifies Taylor-expanding $\epsilon(\mathbf{m},\omega_{\mathbf{m}} + \omega')$ about $\omega' = 0$ in Eq.~(\ref{eqf}), which yields $\frac{\partial \epsilon(\mathbf{m},\omega_{\mathbf{m}})}{\partial \omega} \omega'$ to lowest order. Setting this into Eq.~(\ref{eqf}) and performing the inverse Laplace transform finally yields an amplitude equation for $\tilde{\Phi}(\mathbf{m},t)$
\begin{equation}
%\fl
\frac{\partial \tilde{\Phi}(\mathbf{m},t)}{\partial t}  + \rmi \omega_{\mathbf{m}} \tilde{\Phi}(\mathbf{m},t) =
 \frac{\rmi e N_{\mathrm{bulk}}}{\epsilon_0 k_{\mathbf{m}}^2 (N_{\mathrm{bulk}}  - 1) \frac{\partial \epsilon(\mathbf{m},\omega_{\mathbf{m}})}{\partial \omega}} \sum_{j \in S_{\mathrm{tail}}} \exp(- \rmi \mathbf{k}_{\mathbf{m}} \cdot \mathbf{r}_j).
\label{eqampl}
\end{equation}
The self-consistent dynamics of the potential and of the tail particles is ruled by this equation and by the equation of motion of these particles
\begin{equation}
\ddot{\mathbf{r}}_j = \frac{\rmi e}{L^3 m_\rme} \sum_{\mathbf{n}} \mathbf{k}_{\mathbf{n}} \ \tilde{\Phi}_j(\mathbf{n}) \exp(\rmi \mathbf{k}_{\mathbf{n}} \cdot \mathbf{r}_j).
\label{delrsecwv}
\end{equation}

These two sets of equations generalize to 3 dimensions the self-consistent dynamics defined in Refs \cite{AEE,EEB}. The study of this dynamics enables recovering Vlasovian linear theory with a mechanical understanding (see \cite{SenFest,E013} for a synthetic presentation). In particular, the reason why Landau damping cannot be a damped eigenmode is shown to be rooted deeply in Hamiltonian mechanics~:
a damped eigenmode must exist along with an unstable one, which is going to dominate with probability 1.
Landau damping is recovered as an analogue of van Kampen phase-mixing effect. This phase-mixing in turn plays an essential role in the calculation of Landau instability in order to cancel the damped eigenmode  (section 3.8.3 of Ref.\  \cite{EEB}). The self-consistent dynamics comes with an important bonus~: it brings the information of particle dynamics in parallel with the wave's.
In particular, it reveals that both Landau damping and instability result
from the same synchronization mechanism of particles with waves,
which explains why there is a single formula for the rates of growth and damping \cite{EZE,EEB,Houches}.
This synchronization mechanism was indeed evidenced experimentally \cite{DovEsMa}.
As we stressed in section \ref{secIntro}, this is absent in the Vlasovian description,
and forces textbooks to come up with complementary mechanical models.
The self-consistent dynamics approach enables to assess these models which are not all correct, unfortunately (see section 4.3.1 of Ref.\ \cite{EEB}~; in particular, though initially published with a \textit{caveat}, the surfer model induces in the mind of students the wrong feeling that trapping is involved in Landau effect).
We point out that in Refs \cite{AEE,EEB} the equivalent of Eqs~(\ref{eqampl})-(\ref{delrsecwv}) was obtained without using any smoothing, but by a direct mechanical reduction of degrees of freedom starting with the $N$-body problem.

For the sake of brevity, we do not develop here the full generalization of the analysis in Refs \cite{AEE,EEB}~; it is lengthy, but straightforward. However, since this analysis unifies spontaneous emission with Landau growth and damping, we recall the result ruling the evolution of the amplitude of a Langmuir wave provided by perturbation calculation where the right hand sides of Eqs~(\ref{eqampl})-(\ref{delrsecwv}) are considered as small of order one. This is natural for Eq.~(\ref{eqampl}) since $N_{\mathrm{tail}} \ll N_{\mathrm{bulk}}$, and for Eq.~(\ref{delrsecwv}) if the Langmuir waves have a low amplitude. Let $J(\mathbf{m},t) = \langle \tilde{\Phi}(\mathbf{m},t)\tilde{\Phi}(- \mathbf{m},t) \rangle$, where the average is over the random initial positions of the tail particles (their distribution being spatially uniform).
Then a second order calculation in $\Phi$ yields \begin{equation}
    \frac {\rmd J(\mathbf{m},t)} {\rmd t}
    = 2 \gamma_{\mathbf{m}{\rm L}} J(\mathbf{m},t) + S_{\mathbf{m} \, \mathrm {\rm{spont}} },
   \label{evampfinal}
\end{equation}
where $\gamma_{\mathbf{m}{\rm L}}$ is the Landau growth or damping rate given by
\begin{equation}
  \gamma_{\mathbf{m} {\rm L}} = \alpha_{\mathbf{m}}
  {\frac {\rmd f_{\rm{red}}} {\rmd v}}\left(\frac{\omega_{\mathbf{m}}}{ k_{\mathbf{m}}} ; \mathbf{m}\right)
  \label{SZ122}
\end{equation}
with
\begin{equation}
  \alpha_{\mathbf{m}}
  = \frac{\pi e^2 }{m_\rme \epsilon_0 k_{\mathbf{m}}^2  \frac{\partial \epsilon(\mathbf{m},\omega_{\mathbf{m}})}{\partial \omega}},
 \label{alj}
\end{equation}
and $f_{\rm{red}}$ is the reduced smoothed distribution function $f_{\rm{red}}(v; \mathbf{m}) = \iint f(v \hat{\mathbf{k}}_{\mathbf{m}} + \mathbf{v}_{\bot}) \ \rmd^2 \mathbf{v}_{\bot}$ where $\hat{\mathbf{k}}_{\mathbf{m}}$ is the unit vector along $\mathbf{k}_{\mathbf{m}}$ and $\mathbf{v}_{\bot}$ is the component of the velocity perpendicular to $\mathbf{k}_{\mathbf{m}}$~; $S_{\mathbf{m} \, \mathrm {\rm{spont}} }$ is given by
\begin{equation}
  S_{\mathbf{m} \, \mathrm {\rm{spont}} } = \frac{2 \alpha_{\mathbf{m}}^2}{\pi e^2  k_{\mathbf{m}} n} f_{\rm{red}}(\frac{\omega_{\mathbf{m}}}{ k_{\mathbf{m}}}),
  \label{Spont}
\end{equation}
where $ n=N/L^3$ is the plasma density. $S_{\mathbf{m} \, \mathrm {\rm{spont}} }$ corresponds to the spontaneous emission of waves by particles and induces an exponential relaxation of the waves to the thermal level in the case of Landau damping (the analogue of what was found in \cite{EZE,EEB}).
The second order calculation for the particles yields the diffusion and friction coefficients of the Fokker-Planck equation ruling the tail dynamics.
This equation corresponds to the classical quasilinear result, plus a dynamical friction term mirroring the spontaneous emission of waves by particles, as found in the one-dimensional case in Refs \cite{EZE,EEB}.

An important aspect of the self-consistent dynamics defined
by Eqs~(\ref{eqampl})-(\ref{delrsecwv}) is that it enables to use the modern tools of nonlinear dynamics and chaos available for finite dimensional systems.
Let us consider two examples.
First, the van Kampen phase-mixing effect leading to Landau damping is now a classical result of Vlasovian theory.
However, one may wonder whether nonlinear effects do not destroy these linear modes and the corresponding phase mixing.
Proving the innocuity of nonlinear effects is the equivalent
of deriving a Kolmogorov-Arnold-Moser (KAM) theorem for a continuous system (the Vlasov-Poisson one).
This tour de force partly earned C.\ Villani the 2010 Fields medal \cite{MV}.
The same result for the above finite dimensional self-consistent dynamics requires the standard KAM theorem only~: it is much simpler to keep the genuine granularity of the plasma.

Second, consider a tail distribution function which is a plateau in both velocity and space
(this occurs for instance at the saturation of the bump-on-tail instability
in a particle description of the plasma).
Then the source term in Eq.~(\ref{eqampl}) vanishes, as well as mode coupling,
and the waves keep a fixed amplitude~:
the self-consistency of Eqs~(\ref{eqampl})-(\ref{delrsecwv}) is quenched,
even when particle dynamics is strongly chaotic in the plateau domain.
Then, it is possible to use the tools of 1.5 degree-of-freedom Hamiltonian chaos to compute the diffusion of particle velocities.
In particular, if chaos is strong enough, one may use a quasilinear diffusion coefficient (see section 2.2 of \cite{EE2}).
In a Vlasovian description, the bump-on-tail instability saturates with the previous plateau substituted with a very jagged distribution in both space and velocity resulting from the chaotic stretching and bending of the initial beam-plasma distribution ($f$ is conserved along particle motion)~;
a plateau in velocity exists for the spaced-averaged distribution function only,
and a plateau in space exists for the velocity-averaged distribution only.

%ZZZZZZZZZZZZZZZZZZZZZZZZZZZZZZZZZ
\section{Debye shielding and collisional transport}
\label{DSCT}
%ZZZZZZZZZZZZZZZZZZZZZZZZZZZZZZZZZ

As a further benefit from our many-body approach, this section revisits collisional transport
with the aim of providing a derivation covering all the scales of the impact parameter,
from the classical distance of minimum approach to infinity,
including the scales about the interparticle distance.
For simplicity, we give here the principle of the general derivation
by computing the trace of the diffusion tensor of a given particle.
We perform an explicit mechanical calculation by considering
that particles interact through their shielded Coulomb potentials.

To this end, we focus on the case where the particles have random initial positions,
i.e.\ where the plasma has a uniform density,
and for simplicity we consider the plasma to be in thermal equilibrium.
Then the dynamics of particles has no collective aspect,
but is ruled by the cumulative effect of two-body deflections.
More specifically, we choose random $\mathbf{r}_{l0}$'s,
and vanishing $\delta \mathbf{r}_l(0)$'s and $\delta \dot{\mathbf{r}}_l(0)$'s~;
in contrast to the randomness of initial positions, each particle has a well prescribed initial velocity,
in such a way that the overall initial smoothed velocity distribution is close to some given Maxwellian.
We focus on particle $l$ which is assumed to be close to the center of the cube
with side $L \gg \lambda_{\mathrm{D}}$.
In this section, we approximate the true dynamics with that due to the shielded Coulombian interactions,
i.e.\ we write
\begin{equation}
\delta \ddot{\mathbf{r}}_l = \sum_{j \in S;j \neq l} \mathbf{a}(\mathbf{r}_l-\mathbf{r}_j,\mathbf{v}_j),
\label{delrsecscreen}
\end{equation}
with
\begin{equation}
\mathbf{a}(\mathbf{r},\mathbf{v}) = \frac{e}{m_\rme} \nabla \delta \Phi (\mathbf{r},\mathbf{v}),
\label{acc}
\end{equation}
where $\delta \Phi (\mathbf{r},\mathbf{v})$ is given by Eq.~(\ref{phi}).
This means that we use Eq.\ (\ref{phij})
by substituting $\delta \Phi(\mathbf{r} - \mathbf{r}_j(0) - \dot{\mathbf{r}}_j(0) t,\dot{\mathbf{r}}_j(0))$
with $\delta \Phi(\mathbf{r} - \mathbf{r}_{j},\mathbf{v}_j)$~:
the shielded potential of particle $j$ is computed by taking into account its actual position,
since the genuine shielded potential is the original Coulomb one close to $\mathbf{r}_{j}$.
The error made for $\mathbf{r} - \mathbf{r}_{j} $ of the order of $\lambda_{\rm{D}}$
is small as long as the mismatch of $\mathbf{r}_{j} $ from the ballistic orbit
is much smaller than $\lambda_{\rm{D}}$.
As was done for the bare potential of Eq.\ (\ref{phitildetotM}),
the field acting on a given particle $l$ is obtained
by removing its own divergent contribution $\delta \Phi_l$ from $\Phi$.

We now compute particle $l$ deflection in a sequence of steps. First, we use first order perturbation theory in $\delta \Phi$, which shows the total deflection to be the sum of the individual deflections due to all other particles. For an impact parameter $b$ much smaller than $\lambda_{\rm{D}}$, the deflection due to a particle turns out to be the perturbative value of the Rutherford deflection due to this particle if it were alone. Second, for a close encounter with particle $n$, we show that the deflection of particle $l$ is exactly the one it would undergo if the other $N-2$ particles were absent. Third, the deflection for an impact parameter of order $\lambda_{\rm{D}}$ is shown to be given by the Rutherford expression multiplied by some function of the impact parameter reflecting shielding. These three steps yield an analytical expression for deflection whatever the impact parameter in the ``large box limit'' $L/\lambda_{\mathrm{D}} \rightarrow \infty$.

We first compute $\delta \mathbf{r}_l$ by first order perturbation theory in $\delta \Phi$, taking the ballistic motion defined by Eq.~(\ref{rl0}) as zeroth order approximation. This yields
\begin{equation}
  \delta \dot{\mathbf{r}}_{l1}(t)
  = \sum_{j \in S;j \neq l} \delta \dot{\mathbf{r}}_{lj1}(0,t) ,
\label{delrl1}
\end{equation}
where
\begin{equation}
  \delta \dot{\mathbf{r}}_{lj1}(t_1,t_2)
  = \int_{t_1}^{t_2}  \mathbf{a}[\mathbf{r}_l^{(0)}(t')-\mathbf{r}_j^{(0)}(t'),\mathbf{v}_j] \, \rmd t'.
\label{delrlj1}
\end{equation}
It is convenient to write
\begin{equation}
  \mathbf{r}_l^{(0)}(t')-\mathbf{r}_j^{(0)}(t')
  = \mathbf{b}_{lj} + (t' - t_{lj}) \Delta \mathbf{v}_{lj} ,
  \label{blj}
\end{equation}
where $t_{lj}$ is the time of closest approach of the two ballistic orbits, and $\mathbf{b}_{lj}$ is the vector joining particle $j$ to particle $l$ at this time. Then $b_{lj} = \| \mathbf{b}_{lj}\|$ is the impact parameter of these two orbits when singled out.
The initial random positions of the particles translate into random values of $\mathbf{b}_{lj}$ and of $t_{lj}$.
The typical duration of the deflection of particle $l$ given by Eq.~(\ref{delrlj1}) is $\Delta t_{lj} \equiv b_{lj} / \Delta v_{lj}$
where $\Delta v_{lj} = \| \Delta \mathbf{v}_{lj} \|$, but a certain number, say $\alpha$,
of $\Delta t_{lj}$'s are necessary for the deflection to be mostly completed.
For a given $b_{lj} $ and for $t \gg \Delta t_{lj}$, the deflection of particle $l$ given by Eq.~(\ref{delrlj1}) is maximum if $t_{lj}$ is in the interval $[\alpha \Delta t_{lj}, t - \alpha \Delta t_{lj}]$.
We notice that $\Delta t_{lj}$ is about the inverse of the plasma frequency for $b_{lj} \sim \lambda_{\rm{D}}$ and $\Delta v_{lj}$ on the order of the thermal velocity.

For brevity, we compute here just the trace $T_D$ of the diffusion tensor for the particle velocities.
To this end, we perform an average over all the $\mathbf{r}_{l0}$'s to obtain
\begin{equation}
  \langle \delta \dot{\mathbf{r}}_{l1}^2(t)\rangle
  = \sum_{j \in S;j \neq l}   \langle \delta \dot{\mathbf{r}}_{lj1}^2(t)\rangle,
\label{delrlj1sq}
\end{equation}
taking into account Eq.~(\ref{phi}), and the fact that the initial positions are independently random, as well as the $\mathbf{r}_i - \mathbf{r}_j$'s for $i \neq j$.
Therefore, though being due to the simultaneous scattering of particle $l$ with the many particles inside its Debye sphere, $\langle \delta \dot{\mathbf{r}}_{l1}^2(t)\rangle$ turns out to be the sum of individual two-body deflections for $b_{lj} $'s such that first order perturbation theory is sufficient.
Hence the contribution to $\langle \delta \dot{\mathbf{r}}_{l1}^2(t)\rangle$ of particles with given $b_{lj}$ and $\Delta v_{lj}$ can be computed as if it would result from successive two-body collisions, as was done in Ref.\ \cite{Ros} and in many textbooks.

For an impact parameter much smaller than $\lambda_{\rm{D}}$, the main contribution of $\mathbf{a}[\mathbf{r}_l^{(0)}(t')-\mathbf{r}_j^{(0)}(t'),\mathbf{v}_j]$ to the deflection of particle $l$ comes from times $t'$ for which $\| \mathbf{r}_l^{(0)}(t')-\mathbf{r}_j^{(0)}(t') \| \ll \lambda_{\rm{D}}$.
Therefore $\mathbf{a}(\mathbf{r},\mathbf{v})$ takes on its bare Coulombian value, and $\langle \delta \dot{\mathbf{r}}_{l1}^2(t)\rangle$ is a first order approximation of the effect on particle $l$ of a Rutherford collision with particle $j$.
Comparing this approximate value with the exact one shows the perturbative calculation to be correct for $b_{lj} \gg \lambda_{\rm{ma}} = \frac{e^2}{ \pi m_\rme \epsilon_0 \Delta v_{lj}^2}$,
the distance of minimum approach of two electrons in a Rutherford collision,
as given by energy conservation.

Second, we consider the case of the close approach of particle $p$ to particle $l$,
i.e.\ $b_{ln} \sim \lambda_{\rm{ma}}$. We write the acceleration of particle $l$ as
\begin{equation}
  \ddot{\mathbf{r}}_l
  = \mathbf{a}(\mathbf{r}_l - \mathbf{r}_p, \mathbf{v}_p)
     + \sum_{j \in S;j \neq l,p} \mathbf{a}(\mathbf{r}_l - \mathbf{r}_j, \mathbf{v}_j).
\label{delrsecEx}
\end{equation}
For particle $p$, we write the same equation by exchanging indices $l$ and $p$.
Since the two particles are at distances much smaller
than the inter-particle distance $d = n^{-1/3} = L/N^{1/3}$,
the accelerations imparted on them by all other particles are almost the same.
Therefore, when subtracting the two rigorous equations of motion,
the two summations over $j$ almost cancel.
Moreover, as particles $p$ and $l$ are close,
their shielded potential reduces to the bare Coulomb one, yielding
\begin{equation}
  \frac{\rmd^2 (\mathbf{r}_l - \mathbf{r}_p)}{\rmd t^2}
  = 2 \mathbf{a}_{\rm{C}}(\mathbf{r}_l - \mathbf{r}_p),
\label{delrsecEx2}
\end{equation}
which is the equation describing the Rutherford collision of these two particles in their center of mass frame,
in the absence of all other particles.
Since $b_{lp} \ll d$, $\Delta t_{lp}$ is much smaller than the $\Delta t_{lj}$'s of the other particles.
Therefore the latter produce a negligible deflection of the center of mass during the Rutherford two-body collision,
and the deflection of particle $l$ during this collision is exactly that of a Rutherford two-body collision.
The contribution of such collisions to $\langle \delta \dot{\mathbf{r}}_{l}^2(t)\rangle $
was calculated in Ref.\ \cite{Ros}.

Now, since the deflection of particle $l$ due to particle $j$ as computed by the above perturbation theory
is an approximation of the Rutherford deflection for the same impact parameter,
we may approximate the perturbative deflection with the full Rutherford one,
which provides an obvious matching of the theories for $b_{lj} \sim \lambda_{\rm{ma}}$
and for $\lambda_{\rm{D}} \gg b_{lj} \gg \lambda_{\rm{ma}}$~:
we may use the estimate of \cite{Ros} in the whole domain $b_{lj} \ll \lambda_{\rm{D}}$.

Third, we deal with impact parameters of the order of $\lambda_{\rm{D}}$ and
consider the limit $L / \lambda_{\mathrm{D}} \to \infty$.
Then the deflection due to particle $j$ must be computed with Eq.~(\ref{delrlj1}).
For simplicity, we do the calculation for the case where $\mathbf{v}_j$ is small,
which makes $\delta \Phi (\mathbf{r},\mathbf{v}) \simeq \delta \Phi (\mathbf{r},\mathbf{0})$
which is the Yukawa (or Debye-like) potential $\delta \Phi_{\rm{Y}} (\mathbf{r}) = - \frac{e}{4 \pi \epsilon_0 \| \mathbf{r} \|} \exp (- \frac{\| \mathbf{r} \|}{\lambda_{\rm{D}}})$
(Eq.~(18) of Ref.\ \cite{Gasio}); in this limit $L / \lambda_{\mathrm{D}} \to \infty$.
The first order correction in $\mathbf{k}_{\mathbf{m}}  \cdot \mathbf{v}_j$ to this approximation
is a dipolar potential with an electric dipole moment proportional to $\mathbf{v}_j$.
Since a Maxwellian distribution is symmetrical in $\mathbf{v}$,
these individual dipolar contributions cancel globally.
As a result, the first relevant correction to the Yukawa potential is of second order
in $\mathbf{k}_{\mathbf{m}}  \cdot \mathbf{v}_j$.
This should make the Yukawa approximation relevant for a large part of the bulk of the Maxwellian distribution.

In the small deflection limit, a calculation using the fact that the force derives from a central potential
shows that the full deflection of particle $l$ due to particle $j$ is provided by
\begin{equation}
%\fl
\delta \dot{\mathbf{r}}_{lj1}(- \infty,+ \infty) =
\frac{e^2}{ 4 \pi m_\rme \epsilon_0} \mathbf{b}_{lj}
\int_{- \infty}^{+ \infty}  [\frac{1}{r^3(t)} + \frac{1}{\lambda_{\rm{D}} r^2(t)}] \exp [- \frac{r(t)}{\lambda_{\rm{D}}}]\rmd t,
\label{delrljT}
\end{equation}
where $r(t) = (b_{lj}^2 +\Delta v_{lj}^2 t^2)^{1/2}$ and $\mathbf{b}_{lj}$ was defined with Eq.~(\ref{blj}). Defining $\theta = \arcsin [ \Delta v_{lj} t / r(t)]$, this equation becomes
\begin{equation}
\delta \dot{\mathbf{r}}_{lj1}(- \infty,+ \infty) =
- \frac{2 e^2}{ 4 \pi m_\rme \epsilon_0 \Delta v_{lj}}
 \, \frac{h(b_{lj})}{b_{lj}^2 } \, \mathbf{b}_{lj},
\label{delrljTfin}
\end{equation}
where
\begin{equation}
%\fl
  h(b) =
\int_{0}^{\pi/2} [ \cos (\theta) +  \frac{b}{\lambda_{\rm{D}}}] \exp [- \frac{b}{\lambda_{\rm{D}} \cos (\theta)}] \ \rmd \theta
< [1 + \frac{\pi b}{2 \lambda_{\rm{D}}}] \exp [- \frac{b}{\lambda_{\rm{D}}}].
\label{delrljT2}
\end{equation}
During time $t \gg \Delta t_{lj}$, a volume $2 \pi \Delta v_{lj} t b_{lj} \delta b_{lj}$
of particles with velocity $\mathbf{v}_j$ and impact parameters between $b_{lj}$ and $b_{lj} + \delta b_{lj}$
produce the deflection of particle $l$ given by Eq.~(\ref{delrljTfin}),
and a contribution scaling like $\frac{h^2(b_{lj})}{b_{lj}} \delta b_{lj}$
to $\langle \delta \dot{\mathbf{r}}_{l1}^2(t)\rangle$.
Let $b_{\rm{min}}$ be such that $\lambda_{\rm{D}} \gg b_{\rm{min}} \gg \lambda_{\rm{ma}}$.
The contribution of all impact parameters between $b_{\rm{min}}$ and some $b_{\rm{max}}$
is thus scaling like the integral $\int_{b_{\rm{min}}}^{b_{\rm{max}}} h^2(b)/b  \ \rmd b$.
Since $h(0) \simeq 1$ for $b$ small,
if $b_{\rm{max}} \ll \lambda_{\rm{D}}$ this is the non-shielded contribution
of orbits relevant to the above perturbative calculation.
Since, on approximating it with the Rutherford-like result of Ref.\ \cite{Ros},
this contribution matches that for impact parameters on the order of $\lambda_{\rm{ma}}$,
the contribution of all impact parameters between $\lambda_{\rm{ma}}$
and some $b_{\rm{max}}$ small with respect to $\lambda_{\rm{D}}$
is thus scaling like the integral $\int_{\lambda_{\rm{ma}}}^{b_{\rm{max}}} 1/b  \ \rmd b$
as was computed in Ref.\ \cite{Ros}.
The matching of this result for $b \sim \lambda_{\rm{D}}$
is simply accomplished by setting a factor $h^2(b)$ in the integrand
which makes the integral converge for $b \rightarrow \infty$.
Taking this limit, one finds that the Coulomb logarithm $\ln (\lambda_{\rm{D}} / \lambda_{\rm{ma}})$
of the second Eq.~(14) of Ref.\ \cite{Ros}
becomes $\ln (\lambda_{\rm{D}} / \lambda_{\rm{ma}}) + C$ where $C$ is of order unity.
If the full dependence of the shielding on $\mathbf{v}_j$ were taken into account,
the modification of the Coulomb logarithm would be velocity dependent.

%TTTTTTTTTTTTTTTTTTTTTTTTTTTTTTTTTTTTTT
\section{Conclusion and reflections}
\label{Concl}
%TTTTTTTTTTTTTTTTTTTTTTTTTTTTTTTTTTTTTT

This paper has set new foundations of basic plasma physics by using $N$-body mechanics only.
More specifically, it
%revisited Debye shielding, collisional transport, Landau damping of Langmuir waves,
% and spontaneous emission of these waves. It has
provided a direct path from microscopic mechanics to Debye shielding and Landau damping
without appealing to a lot of extraneous mathematics,
but by using Newton's second law for the $N$-body description, and standard tools of calculus.
The theory has been extended to accommodate a correct description of trapping or chaos due to Langmuir waves,
or to avoid the small amplitude assumption for the electrostatic potential.
Using the shielded potential, collisional transport has been computed for the first time by a convergent expression
including the correct calculation of deflections for all impact parameters.
Shielding and collisional transport have been found to be two related aspects of the repulsive deflections of electrons.

Thanks to its direct approach, this paper also \emph{unifies} Landau growth or damping and spontaneous emission, Debye shielding and collisional transport, and the descriptions of Debye shielding and of linear Langmuir waves waves for both smooth and non-smooth velocity distribution functions.
All these results come with a considerable simplification of the mathematical framework with respect to textbooks and with new intuitive insights into microscopic plasma physics. They might have been derived decades ago, but the present approach worked completely beyond reasonable expectation. In reality, this work is the outcome of a brainstorming about plasma physics \cite{SenFest,E013}, which was first an incentive to revisit collisional transport using shielded potentials. Once this had been done, it looked somewhat odd to use shielded potentials derived by a kinetic approach in a mechanical description with particles. This triggered successively the calculations of sections \ref{FEP}, \ref{SCP}, \ref{STSP} and \ref{MIIDS}, and  Appendix \ref{FNLEP}.
This chaotic research path illustrates Feynman's reflection,
``Perhaps a thing is simple if you can describe it fully in several different ways
without immediately knowing that you are describing the same thing." \cite{Feynman}.

For an expert, it might be hard to feel it useful to simplify the derivation of well-known phenomena in plasma physics. Indeed, for her/him the intricacies underlying such principles are so well assimilated that she/he has difficulties in recognizing them. However, ``difficult'' and ``easy'' have no absolute definition, and our new theory might benefit to students and to their teachers. The former, because of the unification and of the simplification of basic plasma physics brought by the $N$-body approach~; this is all the more important in view of the huge scope of present plasma science.
The latter might gain from compact calculations, proceeding in a continuous way from first principles,
and benefiting from the intuitive nature of mechanics.
This intuitive aspect is important, for it brings a kind of quality insurance when building a course,
even if there is not enough time to teach all the details of the mechanical description of plasmas.
The $N$-body dynamics has always been the ultimate reference in plasma textbooks~:
here it becomes a practical tool. Furthermore, as to chaotic dynamics,
much more is known for finite dimensional systems than for the Vlasov-Poisson system.
It is now possible to avoid the painstaking prerequisites of fluid and kinetic tools,
and to introduce basic plasma concepts with the mechanical approach that reveals their physical content.
Reversing this perspective, the power and flexibility of these tools may now be illustrated
by a recalculation of some basic plasma phenomena.

One might think about trying to apply the above mechanical approach to plasmas with more species, or with a magnetic field, or where particles experience trapping and chaotic dynamics. The first generalization sounds rather trivial, and the third one is under way, at least in one dimension (see a pedestrian introduction in \cite{Houches} and more specific results in \cite{BEEB,BEEBEPS}).
%ion-electron coll: same cross-section classical and meca Q

As in many textbooks, linearization was applied in this paper without questioning deeply its range of validity. However, the smallness of the perturbation is not a sufficient criterion.
Indeed, as reviewed in Ref.\ \cite{HS}, perturbation theory that relies on linearization has to be questioned,
as it yields a solution of the linearized set of equations only.
Whether it also generates a solution of the full set has to be shown explicitly,
and this may be a hard (yet innovative and physically illuminating) task
-- as is for instance the full proof of existence of Landau damping \cite{MV} in a Vlasovian frame,
recalled in section \ref{WPDRT}.

Ph.~Choquard, L.~Cou\"edel, M.-C.~Firpo, W.~Horton, P.K.~Kaw, J.T.~Mendon\c{c}a, F.~Pegoraro, Y.~Peysson,
H.~Schamel, D.~Zarzoso, and J.-Z.~Zhu are thanked for very useful comments and new references.

%VVVVVVVVVVVVVVVVVVVVVVVVVVVVVVVVVVVVVVV
\appendix
%VVVVVVVVVVVVVVVVVVVVVVVVVVVVVVVVVVVVVVV
\section{Discussion of Approximation 1 (corrections to the ballistic approximation and Coulomb potential)}
\label{DA1}
%VVVVVVVVVVVVVVVVVVVVVVVVVVVVVVVVVVVVVVV

Approximation 1 of $\varphi$ by $\phi$ in section \ref{FEP} corresponds to substituting the true dynamics in Eq.\ (\ref{rsectot}) with an approximate one ruled by
\begin{equation}
  \delta \ddot{\mathbf{r}}_l
  = \frac{e}{m_\rme} \nabla \phi_l(\mathbf{r}_l^{(0)} + \delta \mathbf{r}_l),
\label{rsec}
\end{equation}
where $\phi_l (\mathbf{r}) = \sum_{j \in S;j \neq l} \delta \phi_{j} (\mathbf{r})$ is the inverse Fourier transform of Eq.\ (\ref{phitildn}), so that
\begin{equation}
  \lim_{L \to \infty} \delta \phi_{j} (\mathbf{r})
  = - \frac{e}{4 \pi \epsilon_0 \| \mathbf{r} - \mathbf{r}_j^{(0)}\|}
- \frac{e \, \delta \mathbf{r}_j \cdot (\mathbf{r} - \mathbf{r}_j^{(0)})}
    {4 \pi \epsilon_0 \| \mathbf{r} - \mathbf{r}_j^{(0)}\|^3}.
\label{deltaphi}
\end{equation}
The $j$-th contribution to the approximate electric field acting on particle $l$ turns out to be due to a particle located at $\mathbf{r}_{j}^{(0)}$ instead of $\mathbf{r}_j$, and is made up of a Coulombian part and of a dipolar part with dipole moment $- e \, \delta \mathbf{r}_j$.
The cross-over between these two parts occurs for $\| \mathbf{r}_l - \mathbf{r}_j^{(0)}\|$ on the order of $\| \delta \mathbf{r}_j \|$, i.e.\ when the distance between particle $l$ and the ballistic particle $j$ is about the distance between the latter and the true particle $j$. For larger values of $\| \mathbf{r}_l - \mathbf{r}_j^{(0)}\|$, the dipolar component is subdominant. For smaller ones, it is dominant, but with a direction which is a priori random with respect to the Coulombian one ($(\mathbf{r}_l - \mathbf{r}_j^{(0)})$ is almost independent from $\delta \mathbf{r}_j$). Since the $\| \delta \mathbf{r}_j \|$'s are assumed small, the latter case should be rare as it corresponds to a very close encounter between particle $l$ and the ballistic particle $j$. As a result, the approximate electric field stays dominantly of Coulombian nature, but with a small mismatch of the charge positions with respect to the actual ones.

%VVVVVVVVVVVVVVVVVVVVVVVVVVVVVVVVVVVVVVV
\section{Fundamental nonlinear equation for the potential}
\label{FNLEP}
%VVVVVVVVVVVVVVVVVVVVVVVVVVVVVVVVVVVVVVV

 The derivation of the fundamental nonlinear equation for the potential starts as in section \ref{FEP} till the definition of $\delta \mathbf{r}_l$ after Eq. (\ref{rl0}). Equation~(\ref{rsectot}) is equivalent to
\begin{equation}
  \delta \ddot{\mathbf{r}}_l
  = \frac{\rmi e}{L^3 m_\rme} \sum_{\mathbf{n}} \mathbf{k}_{\mathbf{n}} \ \tilde{\varphi}_l(\mathbf{n})
        \exp[\rmi \mathbf{k}_{\mathbf{n}} \cdot (\mathbf{r}_l^{(0)} + \delta \mathbf{r}_l)].
\label{delrsecNL}
\end{equation}
We split $\tilde{\varphi}_l(\mathbf{m})$ as
\begin{equation}
  \tilde{\varphi}_l(\mathbf{m})
  = \tilde{\phi}_l(\mathbf{m}) + \Delta \tilde{\varphi}_l(\mathbf{m})
\label{phitildnapp}
\end{equation}
where $\tilde{\phi}_l (\mathbf{m})$ is given by Eqs (\ref{phitildn})-(\ref{phitildnj}),
and
\begin{equation}
  \Delta \tilde{\varphi}_l(\mathbf{m})
  = - \frac{e}{\epsilon_0 k_{\mathbf{m}}^2} \sum_{j \in S;j \neq l}
       \exp(- \rmi \mathbf{k}_{\mathbf{m}} \cdot \mathbf{r}_{j}^{(0)}) \, R_j(\mathbf{m}),
\label{Deltaphi}
\end{equation}
with
\begin{equation}
  R_j(\mathbf{m})
  = \exp(- \rmi \mathbf{k}_{\mathbf{m}} \cdot \delta \mathbf{r}_j)
     -1 + \rmi \mathbf{k}_{\mathbf{m}} \cdot \delta \mathbf{r}_j ,
\label{Rj}
\end{equation}
which is of order two in $\delta \mathbf{r}_j$.

The Laplace transform of Eq.~(\ref{delrsecNL}) is
\begin{equation}
%\fl
  \omega^2 \delta \hat{\mathbf{r}}_l(\omega)
  = - \frac{\rmi e}{L^3 m_\rme} \sum_{\mathbf{n}} \mathbf{k}_{\mathbf{n}}
        \exp(\rmi \mathbf{k}_{\mathbf{n}} \cdot \mathbf{r}_{l0})
        \ \Psi_l(\widehat{\tilde{\varphi}}_l \ ; \mathbf{n},\omega + \omega_{\mathbf{n},l})
     + \rmi \omega \delta \mathbf{r}_l(0) - \delta \dot{\mathbf{r}}_l(0).
\label{rLaplNL}
\end{equation}
where carets indicate again the Laplace transformed versions of the quantities in Eq.~(\ref{delrsecNL}),
$\omega_{\mathbf{n},l} = \mathbf{k}_{\mathbf{n}} \cdot \mathbf{v}_{l}$ as before,
and the operator $\Psi_l$ acting on a function $g(\mathbf{m},\omega)$ is defined by
\begin{equation}
  \Psi_l(g\ ; \mathbf{n},\cdot) = g(\mathbf{n},\cdot) \ast T_{l}(\mathbf{n},\cdot),
\label{psil}
\end{equation}
where $\cdot$ stands for the frequencies, $ \ast $ is the convolution product in frequency,
and $T_{l}(\mathbf{n},\omega)$ is the Laplace transform
of $\exp(\rmi \mathbf{k}_{\mathbf{n}} \cdot \delta \mathbf{r}_l)$.
The Laplace transform of Eqs~(\ref{phitildnapp})-(\ref{Rj}) yields
\begin{eqnarray}
%\fl  
&&
  k_{\mathbf{m}}^2\widehat{\tilde\varphi}_l(\mathbf{m},\omega)
  \nonumber \\
%\fl  
& = & k_{\mathbf{m}}^2 \widehat{\tilde\phi}_l^{(00)}(\mathbf{m},\omega)
    + \frac{\rmi e}{\epsilon_0} \sum_{j \in S;j \neq l} \exp(- \rmi \mathbf{k}_{\mathbf{m}} \cdot \mathbf{r}_{j0})
        \ [\mathbf{k}_{\mathbf{m}} \cdot \delta \hat{\mathbf{r}}_j(\omega - \omega_{\mathbf{m},j})
            + \rmi \hat{R}_j(\mathbf{m},\omega - \omega_{\mathbf{m},j})],
  \nonumber \\
%\fl 
&&
\label{phihatnNL}
\end{eqnarray}
where $\hat{R}_j(\mathbf{m},\omega)$ is the Laplace transform of $R_j$, and
% $\omega_{\mathbf{m},j} = \mathbf{k}_{\mathbf{m}} \cdot \mathbf{v}_{j}$
% comes from the time dependence of $\mathbf{r}_l^{(0)}$
% in the exponent of Eqs~(\ref{phitildnj})-(\ref{Deltaphi});
$\widehat{\tilde\phi}_l^{(00)}(\mathbf{m},\omega) $ is
the Laplace transform of $\tilde{\phi}_l (\mathbf{m}) $ computed from Eqs~(\ref{phitildn}) and (\ref{phitildnj})
on setting $\delta \mathbf{r}_j =0$ for all $j$'s in the latter.
Substituting the $\delta \hat{\mathbf{r}}_j$'s with their expression Eq.~(\ref{rLaplNL})  yields
\begin{eqnarray}
%\fl  
&&
  k_{\mathbf{m}}^2 \widehat{\tilde\varphi}_l(\mathbf{m},\omega)
  \nonumber\\
%\fl  
&& \quad
  - \frac{e^2}{ L^3 m_\rme \epsilon_0}
  \sum_{\mathbf{n}} \mathbf{k}_{\mathbf{m}} \cdot \mathbf{k}_{\mathbf{n}}
  \ \sum_{j \in S;j \neq l}
     \frac{\Psi_j(\widehat{\tilde\varphi}_j\ ;
                       \mathbf{n},\omega + \omega_{\mathbf{n},j} - \omega_{\mathbf{m},j})}
          {(\omega - \omega_{\mathbf{m},j})^2}
      \exp[\rmi (\mathbf{k}_{\mathbf{n}}-\mathbf{k}_{\mathbf{m}})
           \cdot \mathbf{r}_{j0}]
  \nonumber\\
%\fl  
& = &
  k_{\mathbf{m}}^2 \widehat{\tilde\phi}_l^{(0)}(\mathbf{m},\omega)
  - \frac{e}{\epsilon_0} \sum_{j \in S;j \neq l}
    \exp(- \rmi \mathbf{k}_{\mathbf{m}} \cdot \mathbf{r}_{j0})
    \hat{R}_j(\omega - \omega_{\mathbf{m},j}),
\label{phihatnfNL}
\end{eqnarray}
where $\widehat{\tilde\phi}_l^{(0)}(\mathbf{m},\omega) $ is the Laplace transform
of $\tilde{\phi}_l (\mathbf{m}) $ computed from Eqs~(\ref{phitildn}) and (\ref{phitildnj})
on setting now $\delta \mathbf{r}_j = \delta \mathbf{r}_j(0) + \delta \dot{\mathbf{r}}_j(0) t$ for all $j$'s in the latter.

Summing Eq.~(\ref{phihatnfNL}) over $l = 1,... N$ and dividing by $N-1$, yields
\begin{eqnarray}
%\fl  
&&
  k_{\mathbf{m}}^2 \widehat{\tilde\varphi}(\mathbf{m},\omega)
\nonumber\\
%\fl
&& - \frac{e^2}{ L^3 m_\rme \epsilon_0}
   \sum_{\mathbf{n}} \mathbf{k}_{\mathbf{m}} \cdot \mathbf{k}_{\mathbf{n}}
      \ \sum_{j \in S} \frac{\Psi_j(\widehat{\tilde\varphi}\ ;
                                                \mathbf{n},\omega + \omega_{\mathbf{n},j} - \omega_{\mathbf{m},j})}
                                      {(\omega - \omega_{\mathbf{m},j})^2}
      \exp[\rmi (\mathbf{k}_{\mathbf{n}}-\mathbf{k}_{\mathbf{m}}) \cdot \mathbf{r}_{j0}]
\nonumber\\
%\fl 
&=&  k_{\mathbf{m}}^2 \widehat{\tilde\phi}^{(0)}(\mathbf{m},\omega)
  - \frac{e}{\epsilon_0} \sum_{j \in S} \exp(- \rmi \mathbf{k}_{\mathbf{m}} \cdot \mathbf{r}_{j0})
                  \hat{R}_j(\omega - \omega_{\mathbf{m},j}),
\label{phihatNL}
\end{eqnarray}
where $\widehat{\tilde\phi}^{(0)}(\mathbf{m},\omega) $ is
$\widehat{\tilde\phi}_l^{(0)}(\mathbf{m},\omega) $ complemented by the missing $l$-th term.
Equation (\ref{phihatNL}) is the sought for fundamental nonlinear equation for the potential,
and is a rigorous consequence of Eqs~(\ref{phitildetotM}) and (\ref{rsectot})~: no approximation was made.
Both $\Psi_j(\widehat{\tilde\varphi}\ ; \mathbf{n},\omega + \omega_{\mathbf{n},j} - \omega_{\mathbf{m},j})$
and $\hat{R}_j(\omega - \omega_{\mathbf{m},j})$ are nonlinear in $\delta \mathbf{r}_j$.

Note that in this paper, we use only a very specific part of the fundamental nonlinear equation (\ref{phihatNL})~: the one involving linearization and smoothing. It would be interesting to study the effect of the coupling of Fourier components with both coherent and incoherent effects, in particular, to perform the analysis of section \ref{WPDRT} by substituting $k_{\mathbf{m}}^2 \hat{U}(\mathbf{m},\omega)$ with $- \frac{e}{\epsilon_0} \sum_{j \in S} \exp(- \rmi \mathbf{k}_{\mathbf{m}} \cdot \mathbf{r}_{j0}) \hat{R}_j(\omega - \omega_{\mathbf{m},j})$. The question arises~: is it possible to recover the hole solutions propagating near thermal velocity or slower, which are smooth and nonlinear structures satisfying the full nonlinear Vlasov-Poisson system (Ref.\ \cite{HS} and references therein)~?

%VVVVVVVVVVVVVVVVVVVVVVVVVVVVVVVVVVVVVVV
\section{Discussion of smoothing}
\label{DSmoo}
%VVVVVVVVVVVVVVVVVVVVVVVVVVVVVVVVVVVVVVV

In order to clarify the meaning and validity of smoothing, we rewrite Eq.\ (\ref{phihat}) as
\begin{eqnarray}
%\fl
   \widehat{\tilde{\phi}}(\mathbf{m},\omega) &=&  \frac{\widehat{\tilde{\phi}}^{(0)}(\mathbf{m},\omega)}{\epsilon_\rmd(\mathbf{m},\omega)}
\nonumber\\
%\fl
&& + \frac{e^2}{ L^3 m_\rme \epsilon_0 \epsilon_\rmd(\mathbf{m},\omega)}
 \sum_{\mathbf{l} \neq \mathbf{0}}
      \frac{\mathbf{k}_{\mathbf{m}} \cdot \mathbf{k}_{\mathbf{m} + \mathbf{l}}}{k_{\mathbf{m}}^2}
      \ \sum_{j \in S} \frac{\widehat{\tilde{\phi}}(\mathbf{m} + \mathbf{l},\omega + \omega_{\mathbf{l},j})} {(\omega - \omega_{\mathbf{m},j})^2}
            \exp[\rmi \mathbf{k}_{\mathbf{l}} \cdot \mathbf{r}_{j0}] ,
\nonumber\\
%\fl
&&
\label{phihatsep}
\end{eqnarray}
where
\begin{equation}
  \epsilon_\rmd(\mathbf{m},\omega)
  = 1 - \frac{e^2}{L^3 m_\rme \epsilon_0}
     \sum_{p \in S} \frac{1}{(\omega - \omega_{\mathbf{m},p})^2}
\label{epsdiscr}
\end{equation}
is the discretized version of the classical plasma dielectric function.
Note that Eq.\ (\ref{phihatsep}) has no pole at the $\omega_{\mathbf{m},j}$'s,
since, for each $j$, $\epsilon_\rmd(\mathbf{m},\omega)$ has a pole
canceling exactly the $(\omega - \omega_{\mathbf{m},j})^2$ contribution.

In this paper, we consider smooth distributions that are close to spatially uniform ones.
We now consider discrete analogues of a spatially uniform continuous velocity distribution $f(\mathbf{v})$.
They are special configurations of the $N$-body system,
where particles move on $b$ monokinetic beams,
and where each beam is a simple cubic array of particles.
The elementary cube of any array has its edges along the three orthogonal directions with coordinates $(x,y,z)$,
and the edge length for the $s$-th beam is $L/n_s$ where $n_s$ is an integer.
Therefore, the number of particles of this beam in the elementary cube with volume $L^3$
is $N_s = n_s^3$, and $N = \sum_{s=1}^{b} N_s$. Beam $s$ has a velocity $\mathbf{u}_s$.

The summation over $j$ in Eq.\ (\ref{phihatsep}) can be decomposed
into a summation over the $b$ beams, and for each beam over its particles.
For all particles of beam $s$,
$\omega_{\mathbf{l},j}$ and $\omega_{\mathbf{m},j}$ take on a single value each.
Therefore the summation over the $N_s$ particles bears on
$\exp[\rmi \mathbf{k}_{\mathbf{l}} \cdot \mathbf{r}_{j0}]$ only.
The corresponding sum vanishes unless the three components of $\mathbf{l}$
are on the simple cubic lattice $\mathcal{A}_s = (n_s {\mathbb{Z}})^3$ with mesh length $n_s$~;
then the sum equals $N_s$.
Therefore Eq.\ (\ref{phihatsep}) becomes
\begin{eqnarray}
%\fl
   \widehat{\tilde{\phi}}(\mathbf{m},\omega)
   &=&  \frac{\widehat{\tilde{\phi}}^{(0)}(\mathbf{m},\omega)}{\epsilon_\rmd(\mathbf{m},\omega)}
\nonumber\\
%\fl
  && + \frac{1}{\epsilon_\rmd(\mathbf{m},\omega)}
                \sum_{s =1}^b  \frac{\omega_{{\mathrm{p}} s}^2}{(\omega - \Omega_{\mathbf{m} s})^2}
                \sum_{\mathbf{l} \neq \mathbf{0},\ \mathbf{l} \in \mathcal{A}_s}
                 \frac{\mathbf{k}_{\mathbf{m}} \cdot \mathbf{k}_{\mathbf{m} + \mathbf{l}}}{k_{\mathbf{m}}^2}
                 \ \widehat{\tilde{\phi}}(\mathbf{m} + \mathbf{l},\omega + \Omega_{\mathbf{l} s}),
\label{phihatbeam}
\end{eqnarray}
where $\omega_{{\mathrm{p}} s} = (N_s/N)^{1/2} \omega_{\mathrm{p}}$ is the plasma frequency of beam $s$,
and $\Omega_{\mathbf{m} s} = \mathbf{k}_{\mathbf{m}} \cdot \mathbf{u}_{s}$.
For $\omega = \Omega_{\mathbf{m} s}$, Eq.\ (\ref{phihatbeam}) is understood using
$\lim_{\omega \to \Omega_{\mathbf{m} s}} \epsilon_\rmd(\mathbf{m},\omega) (\omega - \Omega_{\mathbf{m} s})^2
  = - \omega_{{\mathrm{p}} s}^2$.

The first term in the right hand side of Eqs\ (\ref{phihatsep}) and (\ref{phihatbeam}) is the discretized analogue
of the expression of $\widehat{\tilde{\Phi}}(\mathbf{m},\omega)$ provided by Eq.\ (\ref{phihatL}),
and yields an expression for the shielded potential analogous (in Fourier-Laplace representation)
to the summation of the individual potentials
of Eq.~(\ref{phij}) due to the diagonal elements of operator $\mathcal{E}$.
In order to estimate the contribution of the non-diagonal elements in Eq.\ (\ref{phihatbeam}),
we now proceed iteratively~:
in Eq.\ (\ref{phihatbeam}) we substitute
$\widehat{\tilde{\phi}}(\mathbf{m} + \mathbf{l},\omega + \Omega_{\mathbf{l} s})$
with its value provided by
$\widehat{\tilde{\Phi}}(\mathbf{m}+ \mathbf{l},\omega + \Omega_{\mathbf{l} s})$.

According to Eqs (\ref{phij}) and (\ref{phi}),
the shielded potential of particle $j$ involves a summation over $\mathbf{m}$
where $\epsilon(\mathbf{m},\mathbf{k}_{\mathbf{m}} \cdot \mathbf{v})$ stands at the denominator.
For distances on the order of $\lambda_{\rm{D}}$ from particle $j$,
the larger contributions come from $k_{\mathbf{m}} \lesssim \lambda_{\rm{D}}^{-1}$.
In order to prevent the non-diagonal terms
from modifying the smoothed version of the potential at shielding distances, viz.\ large distances,
we must require $\|\mathbf{k}_{\mathbf{l}}\| \gg \lambda_{\rm{D}}^{-1}$.
This implies $2 \pi n_s/L \gg 1/\lambda_{\rm{D}}$,
hence $n \lambda_{\rm{D}}^3 = N (\lambda_{\rm{D}}/L)^3 \gg (2\pi)^{-3}$
since $n_s^3 = N_s < N$. A similar condition is necessary to correctly describe Langmuir waves. Therefore, smoothing is justified provided there are many particles in the Debye sphere.

%TTTTTTTTTTTTTTTTTTTTTTTTTTTTTTTTTTTTTT
%============================================
% \begin{thebibliography}{99}
%============================================

%\section*{References}


\begin{thebibliography}{0}
\expandafter\ifx\csname natexlab\endcsname\relax\def\natexlab#1{#1}\fi
\expandafter\ifx\csname bibnamefont\endcsname\relax
  \def\bibnamefont#1{#1}\fi
\expandafter\ifx\csname bibfnamefont\endcsname\relax
  \def\bibfnamefont#1{#1}\fi
\expandafter\ifx\csname citenamefont\endcsname\relax
  \def\citenamefont#1{#1}\fi
\expandafter\ifx\csname url\endcsname\relax
  \def\url#1{\texttt{#1}}\fi
\expandafter\ifx\csname urlprefix\endcsname\relax\def\urlprefix{URL }\fi
\providecommand{\bibinfo}[2]{#2}
\providecommand{\eprint}[2][]{\url{#2}}

\end{thebibliography}


\begin{thebibliography}{99}
\smallskip

%\numrefs{1}
%
\bibitem{Abe}
  Abe R 1959
  Giant cluster expansion theory and its application to high temperature plasma
  \textit{Progr. Theor. Phys.} \textbf{22} 213--226
%
\bibitem{AEE}
  Antoni M,  Elskens Y and Escande D~F 1998
  {Explicit reduction of $N$-body dynamics to self-consistent particle-wave interaction}
  \textit{Phys. Plasmas} \textbf{5}  841--852
%
\bibitem{Bal}
  Balescu R 1963
  \textit{Statistical mechanics of charged particles}
  (London: Wiley--Interscience)
%
\bibitem{BH}
  Baus M and Hansen J-P 1980
  Statistical mechanics of simple Coulomb systems
  \textit{Phys. Rep.} \textbf{59} 1--94
%
\bibitem{BEEB}
  Besse N, Elskens Y, Escande D~F and  Bertrand P 2011
  Validity of quasilinear theory~: refutations and new numerical confirmation
  \textit{Plasma Phys. Control. Fusion} \textbf{53}  025012 (36 pp).
%
\bibitem{BEEBEPS}
  Besse N, Elskens Y, Escande D~F and  Bertrand P 2011
  On the validity of quasilinear theory
  \textit{Proc. 38th EPS Conference on Controlled Fusion and Plasma Physics, Strasbourg, 2011}, P2.009
  http://ocs.ciemat.es/EPS2011PAP/pdf/P2.009.pdf
%
\bibitem{BoydSan}
  Boyd T~J and Sanderson J 2003
  \textit{The physics of plasmas}
  (Cambridge: Cambridge University press)
%
\bibitem{Dscree}
  Dewar R~L 2010
  The screened field of a test particle
  \textit{In celebration of K C Hines}
  ed McKellar B H J and Amos K
  (Singapore: World Scientific) 47--73,
  and references therein
%
\bibitem{DovEsMa}
  Doveil F, Escande D~F and Macor A 2005
  Experimental observation of nonlinear synchronization due to a single wave
  \textit{Phys. Rev. Lett.} \textbf{94} 085003 (4 pp)
%
\bibitem{EEB}
  Elskens Y and Escande D 2003
  \textit{Microscopic dynamics of plasmas and chaos}
  (Bristol: IoP Publishing)
%
\bibitem{Houches}
  Escande D~F 2010
  {Wave-particle interaction in plasmas~: A qualitative approach}
   \textit{Long-range interacting systems}
  ed Dauxois Th, Ruffo S and Cugliandolo L~F
  (Oxford: Oxford University press) pp 469--506
%
\bibitem{E013}
  Escande D~F 2013
  {How to face the complexity of plasmas ?}
  \textit{From Hamiltonian chaos to complex systems}
  ed Leoncini X and Leonetti M
  (Berlin: Springer) pp 109--157
  http://hal.archives-ouvertes.fr/docs/00/71/74/51/PDF/Complexity\_of\_plasmas\_Escande.pdf
%
\bibitem{SenFest}
  Escande D~F 2013
  Complexity and simplicity of plasmas
  \textit{Preprint}Ê arXiv:1303.4613
%
\bibitem{EE2}
  Escande D and  Elskens Y 2002
  {Proof of quasilinear equations in the chaotic regime of the weak warm beam instability}
  \textit{Phys. Lett.} A \textbf{302}  110--119
%
\bibitem{EZE}
  Escande D~F, Zekri S and Elskens Y 1996
  Intuitive and rigorous microscopic description of spontaneous emission
     and Landau damping of Langmuir waves through classical mechanics
  \textit{Phys. Plasmas} \textbf{3} 3534--3539
%
\bibitem{Feynman}
  Feynman R P 1965
  The development of the space-time view of quantum electrodynamics
  http://www.nobelprize.org/nobel\_prizes/physics/laureates/1965/feynman-lecture.html
%
\bibitem{Gasio}
  Gasiorowicz S, Neuman M and Riddell R~J~Jr 1956
  Dynamics of ionized media
  \textit{Phys. Rev.} \textbf{101} 922--934
%
\bibitem{GoKa}
  Goldenfeld N and Kadanoff L~P 1999
  Simple lessons from complexity
  \textit{Science} \textbf{284} 87--89
%
\bibitem{HW}
  Hazeltine R~D and Waelbroeck F~L 2004
  \textit{The framework of plasma physics}
  (Boulder: Westview Press)
%
\bibitem{Hub}
  Hubbard J 1961
  The friction and diffusion coefficients of the Fokker-Planck equation in a plasma.~II
  \textit{Proc. Roy. Soc. (Lond.)} A \textbf{261} 371--387
%
\bibitem{Kie13}
  Kiessling M~K-H 2013
  to be published
%
\bibitem{MW}
  Malmberg J~H and Wharton C~B 1964
  Collisionless damping of electrostatic plasma waves
  \textit{Phys. Rev. Lett.} \textbf{13}  184--186
%
\bibitem{MJS}
  Montgomery D, Joyce G and Sugihara R 1968
  Inverse third power law for the shielding of test particles
  \textit{Plasma Physics} \textbf{10} 681--687
%
\bibitem{MV}
  Mouhot C and Villani C 2010
  Landau damping
   \textit{J. Math. Phys.} \textbf{51} 015204 (10 pp)
%
\bibitem{Nicholson}
  Nicholson D~R 1983
  \textit{Introduction to plasma theory}
  (New York: Wiley)
%
\bibitem{OWM}
  O'Neil T~M,  Winfrey J~H and  Malmberg J~H 1971
  {Nonlinear interaction of a small cold beam and a plasma}
  \textit{Phys. Fluids} \textbf{14}  1204--1212
%
\bibitem{OLMSS}
  Onishchenko I~N,  Linetski A~R,  Matsiborko N~G, Shapiro V~D and Shevchenko V~I 1970
  {Contribution to the nonlinear theory of excitation
    of a monochromatic plasma wave by an electron beam}
  \textit{ZhETF Pis. Red.} \textbf{12}  407--411
  (Eng. transl. \textit{JETP Lett.} \textbf{12}  281--285)
%
\bibitem{Ros}
  Rosenbluth M~N, MacDonald W~M and Judd D~L 1957
  Fokker-Planck equation for an inverse-square force
  \textit{Phys. Rev.} \textbf{107} 1--6
%
\bibitem{Rost}
  Rostoker N 1964
  Superposition of dressed test particles
  \textit{Phys. Fluids} \textbf{7} 479--490
%
\bibitem{Salp}
  Salpeter E~ÊE 1958
  On Mayer's theory of cluster expansions
  \textit{Ann. Physics} \textbf{5} 183--223
%
\bibitem{HS}
  Schamel H 2012
  Cnoidal electron hole propagation~:
  Trapping, the forgotten nonlinearity in plasma and fluid dynamics
  \textit{Phys. Plasmas} \textbf{19} 020501 (17 pp)
%
\bibitem{Spohn}
  Spohn H 1991
  \textit{Large scale dynamics of interacting particles}
  (Berlin: Springer)
%
\end{thebibliography}
\end{document}